\documentclass[11pt]{article}
\usepackage[utf8]{inputenc}
\usepackage{amsmath,amssymb,amsfonts,bm}
\usepackage[margin=1in]{geometry}
\usepackage{graphicx}
\usepackage{booktabs}
\usepackage{multirow}
\usepackage{array}
\usepackage[hidelinks]{hyperref}
\usepackage{siunitx}
\usepackage{float}
\usepackage{placeins}
\usepackage{tabularx}
\usepackage{longtable}
\usepackage{pdflscape}
\usepackage{caption}
\usepackage{ragged2e}
\usepackage{microtype}
\usepackage{xltabular}

\newcolumntype{Y}{>{\raggedright\arraybackslash}X}

\newcolumntype{P}[1]{>{\RaggedRight\arraybackslash}p{#1}}

\begin{document}

\title{Comparative assessment of germanium-based spin-qubit modalities:\\ donor, acceptor, gate-defined hole, and gate-defined electron platforms}

\author{D.-M.~Mei\thanks{Corresponding author: \texttt{dongming.mei@usd.edu}}, K.-M.~Dong, S.~A.~Panamaldeniya, A.~Prem,\\S.~Chhetri, N.~Budhathoki, and S.~Bhattarai\\[0.4em]Department of Physics, University of South Dakota, Vermillion, SD 57069, USA}

\date{}

\maketitle

\begin{abstract}
High-purity germanium (Ge) has re-emerged as a leading semiconductor platform for spin-based quantum information processing because it combines mature materials processing, access to spin-free isotopes, small carrier effective masses, high mobilities, and strong yet engineerable spin--orbit coupling. At the same time, ``Ge qubits'' do not constitute a single technology: donor spin qubits, acceptor spin qubits, gate-defined hole spin qubits, and gate-defined electron spin qubits exploit different parts of the band structure and therefore make fundamentally different trade-offs among coherence, controllability, fabrication complexity, and scalability. In this work, we present a comparative assessment of these four Ge-based qubit modalities on a common physical and architectural footing. We first review the Ge materials physics that cuts across platforms, including isotopic purification, the multivalley $L$-point conduction band, the spin-$3/2$ valence band, heavy-hole/light-hole mixing, and the roles of strain, interfaces, disorder, and phonons. We also introduce a common framework for estimating the phononic-crystal-modified \(T_1\) in Ge spin systems by combining a calibrated reference relaxation rate, a geometry-specific local strain-density-of-states suppression factor, and parasitic relaxation channels introduced by nanofabrication. We then examine the operating principles, advantages, and limitations of each modality. Donor qubits offer atom-like confinement, strong Stark tunability, and access to hybrid electron--nuclear registers, but are constrained by comparatively strong spin-lattice relaxation. Acceptor qubits provide electrically active spin-$3/2$ physics with unusual quadrupolar and strain-coupled functionality, but remain highly sensitive to microscopic environment and comparatively immature experimentally. Gate-defined electron qubits retain the appeal of spin-$1/2$ encoding, yet inherit the full complexity of Ge's multivalley conduction band and remain underdeveloped. In contrast, gate-defined hole qubits in Ge nanostructures and Ge/SiGe heterostructures currently offer the most compelling combination of all-electrical control, demonstrated multiqubit functionality, and architectural scalability. We conclude that Ge supports a genuinely diverse qubit ecosystem, but that gate-defined hole-spin qubits presently represent the clearest route toward scalable Ge-based quantum processors, while donor, acceptor, and gate-defined electron platforms remain important complementary directions for memory, hybrid, and exploratory architectures.
\end{abstract}

\section{Introduction}

The central challenge in quantum computing is no longer the demonstration of an isolated high-performance qubit, but the realization of a hardware architecture that can scale to fault-tolerant operation. In practice, this requires simultaneously reducing physical error rates below quantum-error-correction thresholds while maintaining long-term stability, sufficient qubit connectivity, and a classical control and readout infrastructure that does not overwhelm the cryostat through wiring density, heat load, power dissipation, and crosstalk \cite{Acharya2025BelowThreshold,vanDijk2019ElectronicInterface}. For this reason, the most promising qubit platform is not simply the one that performs best at the few-qubit level, but the one that offers a credible path to dense integration, low-power control, manufacturable device layouts, and compatibility with cryogenic electronics.

Among the many competing hardware platforms, semiconductor spin qubits offer one of the most compelling routes to meeting this scaling challenge. Their defining advantage is that quantum information is encoded in the spin of a single carrier within a device footprint measured in tens of nanometers rather than micrometers. This naturally supports high integration density, leverages the mature materials and fabrication ecosystem of the semiconductor industry, and offers a realistic pathway toward co-integration with cryogenic control electronics and multiplexed readout. Unlike architectures that rely on physically large qubits or complex room-temperature control infrastructure, semiconductor spin qubits are intrinsically well aligned with the engineering demands of large-scale quantum information processing \cite{Scappucci2021,vanDijk2019ElectronicInterface}. Their long-term appeal therefore lies not only in coherence and gate fidelity, but also in the prospect of building processors using a technology stack already optimized for miniaturization, reproducibility, and large-scale manufacturing.

Within the broader family of semiconductor spin qubits, high-purity germanium (Ge) has emerged as one of the most versatile and attractive material platforms. Historically important in the earliest generations of electronics, Ge is now undergoing a renaissance in quantum science because it combines several attributes that are difficult to obtain simultaneously in a single semiconductor: compatibility with advanced semiconductor processing, the availability of spin-free isotopes, small carrier effective masses, high mobilities, strong but engineerable spin--orbit coupling, and compatibility with heterostructure, donor, and nanofabricated quantum-dot approaches \cite{Scappucci2021}. Ge has become a versatile host for semiconductor spin qubits \cite{Scappucci2021}, superconductor--semiconductor hybrids \cite{Xiang2006GeSiJosephson,Ares2025GeProximitizedQD}, topological-device concepts \cite{Kloeffel2011Helical}, and low-threshold quantum sensing \cite{Mei2025PDU}. These application areas rely on different Ge material stacks and should not be conflated. More importantly, the combined materials toolbox makes Ge unusually well suited to address the central bottleneck of spin-based quantum computing: preserving qubit quality while enabling fast control, compact device layouts, and hardware-efficient interconnects.

An important aspect of Ge is the fundamentally different electronic structure of its conduction and valence bands. In the conduction band, electrons occupy anisotropic $L$ valleys, introducing a nontrivial valley and mass-tensor structure that strongly influences confinement, exchange, and device reproducibility for electron-based qubits. In the valence band, by contrast, the low-energy states arise from the spin-$3/2$ heavy-hole (HH) and light-hole (LH) manifold, giving rise to strong electric-dipole spin resonance (EDSR), large and electrically tunable $g$ tensors, and the possibility of fast all-electrical control \cite{Terrazos2021,Wang2021Optimal}. For hole-spin devices, Ge offers additional advantages directly relevant to scalability: the relevant hole states avoid the valley-degeneracy problem that complicates many electron-spin platforms, holes experience suppressed contact hyperfine coupling because they occupy $p$-like orbitals, and the light in-plane effective mass supports strong electrostatic confinement together with high mobility \cite{Scappucci2021}. Taken together, these features make Ge an unusually rich materials platform for spin-based quantum information processing.

At the same time, ``germanium qubits'' do not represent a single technology. At least four distinct Ge-based spin-qubit modalities can be identified: donor spin qubits, acceptor spin qubits, gate-defined hole spin qubits, and gate-defined electron spin qubits. These platforms exploit different parts of the Ge band structure, are governed by different effective control Hamiltonians, and make fundamentally different trade-offs among coherence, electrical controllability, fabrication precision, and scalability. Donor qubits are atom-like systems in which quantum information is encoded in a donor-bound electron spin or donor-assisted electron--nuclear manifold. Acceptor qubits instead rely on a bound valence hole, inheriting spin-$3/2$ physics, quadrupolar couplings, and pronounced sensitivity to strain and interface symmetry. Gate-defined hole-spin qubits use electrostatically confined holes in Ge nanostructures or quantum wells, enabling fast all-electrical manipulation through strong spin--orbit coupling. Gate-defined electron-spin qubits confine conduction electrons in Ge quantum dots, offering the conceptual simplicity of spin-$1/2$ encoding but inheriting the complexity of the multivalley $L$-point conduction band. These four modalities should therefore be viewed not as minor variants of the same idea, but as physically distinct qubit technologies that happen to share a common semiconductor host.

This distinction is important because the same material parameter can play opposite roles across platforms. Strong spin--orbit coupling, for example, is a liability for donor-electron $T_1$, a resource for hole-spin EDSR, a route to quadrupolar sweet spots in acceptors, and a possible source of variability in large gate-defined arrays. Likewise, the valley structure of the Ge conduction band may complicate donor and gate-defined electron devices while leaving hole-based platforms comparatively simpler. Even the outstanding chemical purity that can now be achieved in Ge does not benefit all modalities in the same way. In addition to isotopic purification, Ge can be zone refined in the laboratory to residual impurity concentrations near $10^{11}\,\mathrm{cm^{-3}}$ (roughly 12N material), and subsequent Czochralski growth can produce detector-grade single crystals with impurity concentrations near $10^{10}\,\mathrm{cm^{-3}}$ (roughly 13N), together with controlled dislocation densities compatible with high-performance devices \cite{Yang2014ZoneRefining,Wang2015USDHPGe,Wang2014Dislocation,Bhattarai2024HPGeGrowth}. This level of materials control suppresses random electrostatic and strain disorder at the source, which is especially important for acceptor-based qubits whose spectra are strongly modified by strain and interface symmetry breaking \cite{AbadilloUriel2016,Zhang2023AcceptorStrain}, but it is also highly relevant for donor, electron, and hole qubits, where reduced background disorder should improve uniformity, coherence, and reproducibility across large arrays \cite{Sigillito2015,Scappucci2021}.

The experimental development of these platforms is also highly uneven. Ge hole-spin qubits have advanced rapidly from the first single-qubit demonstrations to fast two-qubit logic, a four-qubit quantum processor, ultrafast coherent manipulation, sweet-spot operation with substantially improved coherence, coherent spin shuttling through gate-defined quantum dots, and robust local control of a 10-spin array \cite{Watzinger2018,Hendrickx2020Fast,Hendrickx2024Sweet,Wang2022,vanRiggelen2024,John2025}. This trajectory is significant because it shows that Ge is not merely an interesting materials system, but a platform already confronting the real bottlenecks of scale-up: electrically efficient control, multiqubit integration, transport, tunability, and device uniformity. Donor qubits remain attractive because of their atom-like reproducibility, large Stark tunability, and potential for relaxed exchange-coupling length scales relative to silicon, although they are limited by comparatively strong spin-lattice relaxation. Acceptor qubits are intrinsically electric-field-active spin-$3/2$ systems with appealing quadrupolar and interface-tunable physics, but in Ge they remain comparatively immature and highly sensitive to central-cell, strain, and interface details. Gate-defined electron qubits are conceptually attractive because of their simpler spin-$1/2$ encoding, but they remain less experimentally developed and must confront the complexity of Ge's multivalley conduction band.

Another important dimension of the Ge platform is that it is compatible with architectural co-design beyond the qubit itself. The same spin--orbit interaction that enables fast electrical control also enhances sensitivity to charge noise and phonon-mediated relaxation, making decoherence engineering a central issue for Ge-based devices. For this reason, phononic-crystal (PnC) engineering is becoming increasingly attractive in Ge architectures \cite{Smelyanskiy2014DonorGePhononic,Mei2025QST}. By tailoring the acoustic density of states, a PnC can suppress unwanted relaxation channels through phononic bandgaps, protect coherence, and create selected cavity or guided modes that mediate interactions beyond strictly nearest-neighbour exchange. In donor-based Ge systems, this approach has been proposed as a route to suppress one-phonon decay while preserving long-range phonon-mediated spin coupling \cite{Smelyanskiy2014DonorGePhononic}. In Ge hole-spin architectures, recent design studies suggest that phononic crystals could enable decoherence engineering, medium-range coupling, and operation in the 1--4~K regime, where cooling power is far less constrained than at dilution-refrigerator base temperature \cite{Mei2025QST}. Thus, Ge offers not only a strong qubit material, but also a plausible pathway toward the co-design of qubits, couplers, and cryogenic interfaces.

In this paper, we provide a detailed technical comparison of the four principal Ge-based spin-qubit modalities: donor spin qubits, acceptor spin qubits, gate-defined hole spin qubits, and gate-defined electron spin qubits. We first summarize the aspects of Ge materials physics that are relevant across platforms, including isotopic purification, the multivalley $L$-point conduction band, the spin-$3/2$ valence band, heavy-hole/light-hole mixing, and the effects of strain, interface fields, and phonons. We then examine each modality in turn, with particular emphasis on the dominant control mechanisms, decoherence channels, experimental status, and architectural implications. The goal is not merely to catalogue milestones, but to ask a sharper question: which intrinsic features of Ge help or hinder each qubit modality, and what do those trade-offs imply for scalability?

The overall picture that emerges is that gate-defined Ge hole-spin qubits currently offer the strongest combination of experimental maturity, electrical controllability, and architectural scalability. Donor qubits remain compelling for highly tunable atom-like and hybrid electron--nuclear concepts, whereas acceptor and gate-defined electron qubits are better viewed as strategically important but earlier-stage directions whose long-term competitiveness will depend on continued advances in materials control, interface engineering, and qubit reproducibility. By placing donor, acceptor, gate-defined hole, and gate-defined electron qubits on the same physical footing, this comparison aims to clarify which Ge-based technologies are presently best positioned to address the dominant challenge in quantum computing: the transition from elegant few-qubit demonstrations to scalable, fault-tolerant, and hardware-efficient quantum processors.

\section{Ge materials physics relevant to all four platforms}

Before discussing the individual qubit platforms, we summarize several aspects of Ge materials physics that are broadly relevant across all four modalities. Most of this section is a literature-based review of established properties of Ge, including isotopic composition, band-edge structure, strain sensitivity, interface effects, and phonon coupling. However, to facilitate cross-platform comparison, we also introduce several simplified scaling arguments and order-of-magnitude estimates, particularly in Secs.~2.3 and 2.4. These estimates are intended only to provide physical intuition and comparative guidance; they are not presented as new validated materials data, nor as a replacement for detailed atomistic modeling, device-specific simulation, or experiment. Likewise, several of the platform-level implications discussed below represent interpretive synthesis by the authors, built on the cited literature, to clarify how common Ge materials properties affect different qubit modalities in different ways.

\subsection{Isotopic composition and hyperfine environment}

A key attraction of Ge is that its host nuclear environment is unusually simple and highly tunable. Natural Ge contains five stable isotopes, \(^{70}\)Ge, \(^{72}\)Ge, \(^{73}\)Ge, \(^{74}\)Ge, and \(^{76}\)Ge, with approximate natural abundances of \(20.5\%\), \(27.4\%\), \(7.8\%\), \(36.5\%\), and \(7.8\%\), respectively; among these, only \(^{73}\)Ge carries nuclear spin, with \(I=9/2\) \cite{NISTGeIsotopes,Scappucci2021}. Thus, unlike III--V compounds, natural Ge already provides a host in which more than \(92\%\) of nuclei are spin free, and isotopic purification can further deplete the residual \(^{73}\)Ge bath in close analogy to isotopically enriched silicon \cite{Scappucci2021,Sigillito2015}. This is an important systems-level advantage because it allows the magnetic-noise environment to be engineered at the materials stage, rather than mitigated only by control protocols or device design.

The consequences of this isotopic structure depend strongly on the qubit modality. For donor-electron qubits, the electronic wavefunction overlaps directly with host nuclei, so residual \(^{73}\)Ge can produce spectral diffusion and dominate dephasing in natural or only partially enriched material \cite{Sigillito2015}. Indeed, pulsed-ESR measurements on donors in Ge have shown that in samples with significant \(^{73}\)Ge content, the coherence time is limited by the nuclear-spin bath, whereas in more strongly enriched material the coherence can become limited instead by spin-lattice relaxation, \(T_2 \simeq 2T_1\) \cite{Sigillito2015}. Gate-defined electron-spin qubits inherit a related sensitivity, although the exact strength of the hyperfine coupling depends on the orbital confinement, valley composition, and the extent to which the wavefunction penetrates barriers or interfaces. In Ge/SiGe heterostructures, the relevant magnetic environment may therefore include not only residual \(^{73}\)Ge in the well but also any magnetic isotopes present in the SiGe barriers, unless the full stack is isotopically engineered \cite{Scappucci2021}.

For hole-based qubits, the situation is more favorable but also more subtle. Because valence-band states have predominantly \(p\)-like symmetry, the Fermi-contact hyperfine term is strongly suppressed relative to conduction electrons, so the residual host nuclear bath is generally less destructive for hole spins than for electron spins \cite{Scappucci2021,Wang2021Optimal,Hendrickx2024Sweet,Stehouwer2025}. This is one reason Ge hole-spin qubits can combine fast all-electrical control with relatively long coherence. However, the hyperfine interaction is not eliminated entirely: anisotropic and dipolar terms remain, and in extended arrays even a weak residual nuclear bath can still contribute to device-to-device variability, frequency diffusion, and slow calibration drift. For acceptor qubits, a bound valence hole likewise benefits from the suppressed contact interaction with the Ge host, but an additional local hyperfine channel can arise from the acceptor dopant itself if the impurity nucleus carries nonzero spin. As a result, the isotopic simplicity of the Ge host is a major advantage for all four qubit classes, but it is most transformative for donor and gate-defined electron qubits, while for hole and acceptor qubits it acts in concert with the orbital symmetry of the valence band to produce an especially favorable magnetic environment.

The dopant nuclear spin is a design-dependent resource or noise channel rather than an intrinsic advantage or disadvantage of one impurity modality. For a donor qubit, an isotope with a controllable nuclear spin can form a long-lived electron--nuclear register, whereas a spin-zero donor or an ionized donor state may be preferred when nuclear coupling is unwanted. For an acceptor qubit, an uncontrolled dopant nuclear spin can add local hyperfine noise to an electrically and strain-sensitive bound-hole state; however, a spin-zero acceptor isotope removes this channel, and a deliberately addressable acceptor nucleus could in principle provide a hybrid register. The comparison therefore depends on isotope choice and control capability.

\subsection{Band-edge structure, effective masses, and implications for Ge spin qubits}

A central reason Ge is such a rich spin-qubit material is that its conduction- and valence-band edges occur at fundamentally different points in the Brillouin zone and therefore generate very different qubit physics. Ge is an indirect-gap semiconductor whose lowest conduction-band minima lie at the four equivalent \(L\) points along the \(\langle 111\rangle\) directions, while the valence-band maximum lies at the zone center \(\Gamma\) \cite{Scappucci2021,IoffeGeParameters,Neamen2012}. At room temperature, the indirect gap associated with the \(L\)-valley minimum is \(E_{g,\mathrm{ind}}\approx 0.66~\mathrm{eV}\), whereas the direct \(\Gamma\)-point conduction minimum lies higher at \(E_{g,\Gamma}\approx 0.80~\mathrm{eV}\) \cite{IoffeGeParameters,Neamen2012}. This distinction is central to Ge spin qubits: donor-electron and gate-defined electron qubits inherit the multivalley \(L\)-point conduction-band structure, whereas acceptor and gate-defined hole qubits inherit the \(\Gamma\)-point valence-band structure and are therefore valley free.

For conduction-band electrons, each \(L\) valley is approximately ellipsoidal, with longitudinal and transverse masses \(m_l\) and \(m_t\). In the anisotropic effective-mass approximation, the characteristic single-valley envelope radii scale schematically as \(a_{\perp}\propto \epsilon_{\mathrm{Ge}}/m_t\) and \(a_{\parallel}\propto \epsilon_{\mathrm{Ge}}/m_l\), where \(\epsilon_{\mathrm{Ge}}\) is the static dielectric constant of Ge. The large mass ratio \(m_l/m_t\) therefore produces an ellipsoidal envelope with a much larger transverse than longitudinal radius. The overall spatial extent is set jointly by dielectric screening and the effective masses; mass anisotropy determines the directional shape rather than a single isotropic radius. Strain or confinement additionally shifts the four \(L\)-valley energies and changes their admixture, thereby modifying the valley--orbit spectrum, \(g\) tensor, and exchange coupling \cite{Li2012GeSpinLifetime,Neamen2012,Pica2016,Baron2003Lvalleyg,Virgilio2009GeValley}.

The valence band of Ge is qualitatively different and, in many ways, more favorable for electrically controlled spin qubits. In bulk Ge, the valence-band maximum at \(\Gamma\) consists of degenerate heavy-hole (HH) and light-hole (LH) bands together with a spin-orbit split-off band separated by \(\Delta_{\mathrm{SO}}\approx 0.29~\mathrm{eV}\) \cite{IoffeGeParameters,Neamen2012}. Representative bulk effective masses are \(m_{hh}\approx 0.28\,m_0\), \(m_{lh}\approx 0.044\,m_0\), and \(m_{so}\approx 0.084\,m_0\) \cite{Neamen2012}. In strained Ge/SiGe quantum wells, confinement and compressive strain lift the HH--LH degeneracy and usually leave a predominantly HH-like ground state, but with substantial residual HH--LH mixing. This mixing is not merely a perturbation; it is the microscopic origin of the strong and electrically tunable spin--orbit interaction that enables EDSR, anisotropic \(g\)-tensors, and sweet-spot operation in Ge hole-spin qubits \cite{Terrazos2021,Wang2021Optimal,Hendrickx2024Sweet}. Moreover, the top valence subband in strained Ge/SiGe can exhibit a very light in-plane effective mass, on the order of \(0.05\,m_0\), which improves tunnel coupling, increases orbital spacing, and relaxes lithographic constraints in dense arrays \cite{Lodari2019LightMass,Scappucci2021}.

It is also useful to distinguish between directional band masses and the averaged effective masses relevant for materials and transport. Because the \(L\)-valley conduction band is anisotropic and fourfold degenerate, the density-of-states and conductivity effective masses differ substantially from the longitudinal and transverse masses of a single valley. Likewise, in bulk Ge, the combined HH and LH contributions define density-of-states and conductivity masses that differ from the individual band-edge masses.

The bulk density-of-states and conductivity masses are included as common reference quantities for carrier statistics, screening, transport, and electrostatic response in bulk Ge and in reservoirs or leads associated with Ge devices. They are therefore useful for comparing materials and device environments across the four modalities. They are not the directional confinement masses of a specific donor envelope and must not be inserted as the quantum-dot mass in a strained Ge/SiGe heterostructure, where the curvature and HH/LH composition of the occupied subband must be used. For the strained quantum-well devices discussed below, a top-subband in-plane mass of order \(0.05m_0\), with a value dependent on stack and gate field, is therefore more directly relevant than the bulk combined masses in Table~\ref{tab:Ge_derived_masses}.

Taken together, these band-structure features strongly shape the relative prospects of the four Ge spin-qubit modalities discussed in this paper. Donor and gate-defined electron qubits benefit from mature electron-spin concepts and the possibility of long coherence in highly purified Ge, but they must manage multivalley \(L\)-point physics and strong mass anisotropy. By contrast, gate-defined hole and acceptor qubits are built from a valley-free \(\Gamma\)-point valence band whose spin-\(3/2\) structure naturally supports strong electrical control. This contrast between the \(L\)-point conduction band and the \(\Gamma\)-point valence band explains much of the emerging hierarchy among Ge qubit platforms.

\begin{table}[htbp!]
\centering
\caption{Conduction-band properties of Ge relevant to electron-based spin qubits.}
\label{tab:Ge_conduction}
\begin{tabular}{|p{2.8cm}|p{2.0cm}|p{3.0cm}|p{6.0cm}|}
\hline
\textbf{Platform} & \textbf{Band edge} & \textbf{Representative masses} & \textbf{Impact on qubit performance} \\
\hline
Donor-spin qubits & \(L\) valleys (4 equivalent) & \(m_l \approx 1.6\,m_0,\; m_t \approx 0.08\,m_0\) & The extended transverse component of the anisotropic donor envelope can ease some spacing constraints, but direction-dependent overlap and multivalley physics complicate exchange, valley--orbit structure, and sensitivity to strain and disorder. \\
\hline
Gate-defined electron qubits & \(L\) valleys (4 equivalent) & \(m_l \approx 1.6\,m_0,\; m_t \approx 0.08\,m_0\) & Valley anisotropy and valley-orbit coupling complicate confinement and spectral uniformity; interfaces and strain can strongly perturb the low-energy spectrum. \\
\hline
\end{tabular}
\end{table}

\begin{table}[htbp!]
\centering
\caption{Valence-band properties of Ge relevant to hole-based spin qubits. Hole masses are strongly strain- and confinement-dependent.}
\label{tab:Ge_valence}
\begin{tabular}{|p{2.8cm}|p{2.0cm}|p{3.0cm}|p{6.0cm}|}
\hline
\textbf{Platform} & \textbf{Band edge} & \textbf{Representative masses} & \textbf{Impact on qubit performance} \\
\hline
Gate-defined hole qubits & \(\Gamma\) point; HH/LH manifold & Top subband in-plane mass can be \(\sim 0.05\,m_0\) in strained Ge/SiGe & Valley-free band edge avoids valley complications; light in-plane mass improves tunnel coupling and orbital spacing; HH--LH mixing enables strong EDSR, tunable \(g\) tensors, and sweet-spot operation. \\
\hline
Acceptor-spin qubits & \(\Gamma\)-derived bound-hole states & No single simple band mass for localized acceptor states & Inherits valley-free spin-\(3/2\) physics, enabling quadrupolar control and strain/electric-field tunability, but is highly sensitive to symmetry breaking and strain disorder. \\
\hline
\end{tabular}
\end{table}

Tables~\ref{tab:Ge_conduction} and \ref{tab:Ge_valence} summarize this contrast at the platform level: Table~\ref{tab:Ge_conduction} highlights how the \(L\)-valley conduction band shapes donor and gate-defined electron qubits, whereas Table~\ref{tab:Ge_valence} highlights how the \(\Gamma\)-point valence band shapes gate-defined hole and acceptor qubits.

The same physics can also be summarized more compactly at the band-structure level. Table~\ref{tab:Ge_band_edges} lists the relevant extrema in \(k\)-space, their energies, and representative band-edge masses, while Table~\ref{tab:Ge_derived_masses} lists the corresponding density-of-states and conductivity masses. Together, these tables bridge the underlying electronic structure of Ge to the practical behavior of spin-qubit devices.

\begin{table}[htbp!]
\centering
\small
\caption{Band-edge structure and representative effective masses in Ge relevant to spin qubits.}
\label{tab:Ge_band_edges}
\begin{tabular}{|p{2.2cm}|p{1.2cm}|p{1.5cm}|p{3.0cm}|p{5.8cm}|}
\hline
\textbf{Band / carrier} & \textbf{k-space location} & \textbf{Energy} & \textbf{Representative mass} & \textbf{Spin-qubit implication} \\
\hline
Direct conduction band & \(\Gamma\) & \(E_{g,\Gamma}\approx 0.80\) eV & \(m^* \approx 0.041\,m_0\) & Higher-energy direct valley; mainly relevant for optical processes and intervalley band-structure considerations rather than low-energy electron qubits. \\
\hline
Indirect conduction band & \(L\) (4 valleys) & \(E_{g,\mathrm{ind}}\approx 0.66\) eV & \(m_l \approx 1.64\,m_0,\; m_t \approx 0.082\,m_0\) & Governs donor and gate-defined electron qubits; multivalley structure and mass anisotropy affect exchange, confinement, and strain sensitivity. \\
\hline
Heavy-hole valence band & \(\Gamma\) & \(E_v = 0\) & \(m_{hh}\approx 0.28\,m_0\) & Part of the spin-\(3/2\) HH/LH manifold; important for acceptor and hole-spin qubits. \\
\hline
Light-hole valence band & \(\Gamma\) & \(E_v = 0\) & \(m_{lh}\approx 0.044\,m_0\) & Light mass and HH--LH mixing enable strong EDSR and fast electrical control. \\
\hline
Split-off valence band & \(\Gamma\) & \(\Delta_{\mathrm{SO}}\approx 0.29\) eV below \(E_v\) & \(m_{so}\approx 0.084\,m_0\) & Usually remote from low-energy qubit dynamics, but important in realistic band-structure modeling. \\
\hline
\end{tabular}
\end{table}

\begin{table}[htbp!]
\centering
\small
\caption{Bulk-Ge density-of-states and conductivity masses. These values provide reference scales for bulk carrier statistics, screening, and transport, including the material surrounding impurity qubits and the reservoirs coupled to gate-defined devices. They are not confinement masses for strained Ge/SiGe quantum wells or device-specific quantum dots.}
\label{tab:Ge_derived_masses}
\begin{tabular}{|p{2.2cm}|p{4.6cm}|p{1.5cm}|p{5.8cm}|}
\hline
\textbf{Quantity} & \textbf{Expression} & \textbf{Value} & \textbf{Physical meaning} \\
\hline
Electron DOS mass & \(m_{d,e}^{*}=N_v^{2/3}(m_l m_t^2)^{1/3}\), with \(N_v=4\) & \(0.56\,m_0\) & Sets the conduction-band density of states; reflects both valley degeneracy and ellipsoidal anisotropy. \\
\hline
Hole DOS mass & \(m_{d,h}^{*}=\left(m_{hh}^{3/2}+m_{lh}^{3/2}\right)^{2/3}\) & \(0.29\,m_0\) & Sets the valence-band density of states from the combined HH and LH bands. \\
\hline
Electron conductivity mass & \(m_{c,e}^{*}= \displaystyle \frac{3}{\frac{1}{m_l}+\frac{2}{m_t}}\) & \(0.12\,m_0\) & Governs transport and electrostatic response after averaging over equivalent electron directions. \\
\hline
Hole conductivity mass & \(m_{c,h}^{*}= \displaystyle \frac{m_{hh}^{3/2}+m_{lh}^{3/2}}{m_{hh}^{1/2}+m_{lh}^{1/2}}\) & \(0.21\,m_0\) & Governs transport after combining HH and LH contributions in the valence band. \\
\hline
\end{tabular}
\end{table}

\subsection{Strain, interfaces, and disorder}

Strain, interfaces, and disorder are not secondary materials issues in Ge spin qubits; they are central design parameters that directly shape qubit coherence, tunability, and reproducibility. In planar Ge/SiGe heterostructures, disorder arises from several coupled sources, including interface roughness, random-alloy fluctuations in the SiGe barriers, remote charged defects, gate-induced electric-field inhomogeneity, and non-uniform strain. Together, these effects modify charge noise, spin-orbit coupling, \(g\)-tensor anisotropy, and qubit-frequency variability, and therefore play a major role in the scalability of Ge hole-spin qubits \cite{Scappucci2021,Stehouwer2025,Sarkar2025}. In particular, random-alloy disorder and gate-induced strain can generate effective spin-orbit terms comparable to, or even larger than, the nominal bulk contributions, while high-quality strained Ge/SiGe quantum wells grown on carefully prepared Ge wafers can significantly reduce charge noise and improve wafer-scale uniformity \cite{Stehouwer2025,Sarkar2025}. At the same time, recent work on hole qubits in unstrained Ge layers suggests that partially removing or re-engineering heterostructure strain may reduce the extreme \(g\)-factor anisotropy of heavy-hole states and thereby simplify field optimization and large-scale device design \cite{Mauro2025}.

For impurity-based qubits, interfaces influence the qubit in a qualitatively different way. In donor and acceptor systems, an interface does not merely add noise; it reshapes the impurity-bound wavefunction itself. The resulting confinement, dielectric mismatch, and broken symmetry can split Kramers doublets, modify selection rules, and enable electric control through parity mixing \cite{AbadilloUriel2016}. This is particularly important for acceptor qubits because the spin-\(3/2\) valence manifold is intrinsically sensitive to local symmetry breaking and strain. Thus, whereas gate-defined hole qubits often exploit strain and interface asymmetry as useful resources for strong electrical control, acceptor qubits can be much more vulnerable to uncontrolled background strain and interface-induced level reshaping.

To support the cross-platform comparison developed later in this paper, we now introduce a deliberately simplified set of order-of-magnitude estimates for impurity-induced strain backgrounds in Ge. These estimates are not intended as precision predictions for a specific crystal-growth or device environment. Rather, they provide a compact comparative framework for thinking about how residual impurity concentration and impurity species can influence strain-sensitive qubit modalities, especially acceptor and hole-based systems.

These observations motivate a useful distinction between \emph{heterostructure-induced strain} and \emph{residual-impurity-induced strain}. The former is typically intentional and relatively large, often in the range \(10^{-3}\) to \(10^{-2}\) in strained Ge/SiGe quantum wells, where it is used to lift HH/LH degeneracy and engineer the valence-band structure \cite{Terrazos2021,Mauro2025}. The latter is far smaller, but becomes increasingly important when the goal is qubit-to-qubit uniformity, ultralow disorder, and the realization of highly strain-sensitive qubit modalities such as acceptor qubits. In high-purity bulk Ge, the average hydrostatic strain induced by dilute substitutional impurities can be estimated using a defect-relaxation-volume picture.

For a substitutional impurity of species \(i\), the relaxation volume \(\Delta\Omega_i\) is the change in crystal volume produced by introducing the defect and allowing the surrounding lattice to relax at zero external stress, relative to the corresponding perfect-crystal reference under the same thermodynamic convention. Equivalently, in linear elasticity it is the spatial integral of the defect-induced dilatation, \(\Delta\Omega_i=\int \mathrm{Tr}[\varepsilon_i(\mathbf r)]\,d^3r\), outside the unresolved atomistic core. We take \(\Delta\Omega_i>0\) for a defect that expands the lattice and \(\Delta\Omega_i<0\) for one that contracts it. This integrated quantity controls the coarse-grained average strain; it should not be confused with the much larger, position-dependent strain close to the impurity.

If impurity species \(i\) changes the local atomic volume by \(\Delta \Omega_i\), then the average volumetric strain is
\begin{equation}
\left\langle \mathrm{Tr}\,\varepsilon \right\rangle_i \simeq c_i \frac{\Delta \Omega_i}{\Omega_0},
\end{equation}
where \(c_i=n_i/N_0\) is the atomic fraction of impurity \(i\), \(n_i\) is its number density, and \(N_0\) is the atomic density of Ge. For diamond-cubic Ge,
\begin{equation}
N_0=\frac{8}{a_0^3}\approx 4.4\times 10^{22}\ \mathrm{cm^{-3}},
\end{equation}
with \(a_0 \approx 5.658\)~\AA\ \cite{IoffeGeBasic} and \(\Omega_0=1/N_0\).

For \(N_i=n_iV\) uncorrelated substitutional impurities of species \(i\) in a coarse-graining volume \(V\), the mean total relaxation volume is \(N_i\Delta\Omega_i\). Dividing by \(V\) gives \(\langle\mathrm{Tr}\,\varepsilon\rangle_i=n_i\Delta\Omega_i\), or equivalently \(c_i\Delta\Omega_i/\Omega_0\). For approximately isotropic dilation, the mean linear strain is one third of the volumetric strain. This dilute, linear-elastic construction is a coarse-grained estimate, not the peak strain close to an individual defect \cite{Cordero2008,Xu2016,Chang2025}.

If the relaxation is approximately isotropic, the corresponding average linear strain is
\begin{equation}
\bar{\varepsilon}_i \simeq \frac{1}{3}\left\langle \mathrm{Tr}\,\varepsilon \right\rangle_i .
\end{equation}

In the ultradilute limit relevant to residual impurities, a convenient order-of-magnitude estimate is obtained by relating the relaxation volume to the covalent-radius mismatch,
\begin{equation}
\frac{\Delta \Omega_i}{\Omega_0}\approx 3\,\eta_i,
\qquad
\eta_i \equiv \frac{r_i-r_{\mathrm{Ge}}}{r_{\mathrm{Ge}}},
\end{equation}
which leads to the simple expression
\begin{equation}
\bar{\varepsilon}_i \approx \eta_i\,c_i
= \eta_i \frac{n_i}{N_0}.
\label{eq:avgstrain}
\end{equation}
Here \(r_i\) is the covalent radius of impurity \(i\), and \(r_{\mathrm{Ge}}\) is the covalent radius of Ge \cite{Cordero2008}. Although approximate, Eq.~(\ref{eq:avgstrain}) is consistent with the expected linear Vegard-like scaling for dilute dopants in Ge and is sufficiently simple to compare different impurity species and purity grades \cite{Xu2016,Chang2025}. For multiple independent impurity species, the total average hydrostatic strain can be estimated as
\begin{equation}
\bar{\varepsilon}_{\mathrm{tot}} \approx \sum_i \eta_i \frac{n_i}{N_0}.
\end{equation}

For qubit variability, however, the mean strain is only part of the story. What often matters more is the spatially averaged \emph{fluctuation} of strain over the finite volume sampled by the qubit wavefunction. If impurities are randomly distributed, the root-mean-square strain fluctuation over a volume \(V\) scales as
\begin{equation}
\varepsilon_{\mathrm{rms}}(V)
\approx
\left[
\sum_i \eta_i^2 \frac{n_i}{N_0^2 V}
\right]^{1/2}.
\label{eq:rmsstrain}
\end{equation}
This scaling shows explicitly that the strain background seen by a mesoscopic qubit decreases both with improved chemical purity and with the spatial averaging associated with the qubit extent. This is one of the main reasons detector-grade Ge is qualitatively different from ordinary electronic-grade Ge for strain-sensitive qubits.

For a Poisson distribution, \(\mathrm{Var}(N_i)=N_i=n_iV\). The fluctuation in impurity fraction is therefore \(\sqrt{n_iV}/(N_0V)\); multiplying by \(\eta_i\) and adding independent species in quadrature yields Eq.~\eqref{eq:rmsstrain}. This coarse-grained result requires \(V\) to contain many possible lattice sites and neglects impurity correlations, clustering, dislocations, interfaces, elastic anisotropy, charge-state-dependent relaxation volumes, and the much larger local strain near an individual defect.

Table~\ref{tab:Ge_impurity_strain} summarizes representative estimates from Eq.~(\ref{eq:avgstrain}) for several common donor and acceptor impurities in Ge. For 5N and 9N material, we use the nominal atomic fractions \(c=10^{-5}\) and \(10^{-9}\), respectively. For detector-grade ``13N'' material, we adopt the practical HPGe convention used in this paper, namely a residual net electrically active impurity density \(n_{\mathrm{imp}}\sim 10^{10}\,\mathrm{cm^{-3}}\), rather than a total chemical impurity concentration summed over all elements. This is the regime targeted and discussed in the USD HPGe growth program~\cite{Yang2015ZoneRefining,Wang2015USDHPGe,Bhattarai2024HPGeGrowth}. Accordingly, the values in Table~\ref{tab:Ge_impurity_strain} should be interpreted as order-of-magnitude background estimates for comparative intuition, not as precision materials benchmarks for a specific boule or device stack.

\begin{table}[htp!]
\centering
\small
\caption{Order-of-magnitude estimates of average impurity-induced linear strain in Ge from Eq.~(\ref{eq:avgstrain}). Covalent radii are from Ref.~\cite{Cordero2008}. Here 5N and 9N are treated as nominal atomic fractions \(10^{-5}\) and \(10^{-9}\), respectively. The label ``13N'' is used in the detector-grade HPGe convention adopted in this paper, meaning a net electrically active impurity density of \(n_{\mathrm{imp}}\approx 10^{10}\,\mathrm{cm^{-3}}\); it should not be interpreted as the total chemical impurity concentration summed over all elements. These values represent average background strains for comparative intuition, not peak local strains near an impurity atom or precision predictions for a specific crystal.}
\label{tab:Ge_impurity_strain}
\begin{tabular}{|c|c|c|c|c|c|c|}
\hline
Impurity & Type & $r_i$ (pm) & $\eta_i=(r_i-r_{\mathrm{Ge}})/r_{\mathrm{Ge}}$ & $|\bar{\varepsilon}_i|$ (5N) & $|\bar{\varepsilon}_i|$ (9N) & $|\bar{\varepsilon}_i|$ (13N) \\
\hline
B  & acceptor & 84  & $-0.300$   & $3.0\times 10^{-6}$  & $3.0\times 10^{-10}$ & $6.8\times 10^{-14}$ \\
Al & acceptor & 121 & $+8.3\times10^{-3}$ & $8.3\times 10^{-8}$  & $8.3\times 10^{-12}$ & $1.9\times 10^{-15}$ \\
Ga & acceptor & 122 & $+1.67\times10^{-2}$ & $1.7\times 10^{-7}$  & $1.7\times 10^{-11}$ & $3.8\times 10^{-15}$ \\
P  & donor    & 107 & $-0.108$   & $1.1\times 10^{-6}$  & $1.1\times 10^{-10}$ & $2.5\times 10^{-14}$ \\
As & donor    & 119 & $-8.3\times10^{-3}$ & $8.3\times 10^{-8}$  & $8.3\times 10^{-12}$ & $1.9\times 10^{-15}$ \\
Sb & donor    & 139 & $+0.158$   & $1.6\times 10^{-6}$  & $1.6\times 10^{-10}$ & $3.6\times 10^{-14}$ \\
\hline
\end{tabular}
\end{table}

Even with this caveat, the comparison is informative. The estimated average impurity-induced strain background drops by roughly four orders of magnitude from 5N to 9N material, and by another three to four orders of magnitude when moving from 9N material to detector-grade 13N Ge. For strongly size-mismatched common impurities, such as B and Sb, the estimated average strain decreases from the \(10^{-6}\) level in 5N material to the \(10^{-10}\) level in 9N material and to the \(10^{-14}\) level in detector-grade 13N Ge. Thus, the purity trend is robust even if the covalent-radius-mismatch model is regarded only as a comparative approximation.

\begin{figure}[htp!!!]
    \centering
    \includegraphics[width=\linewidth]{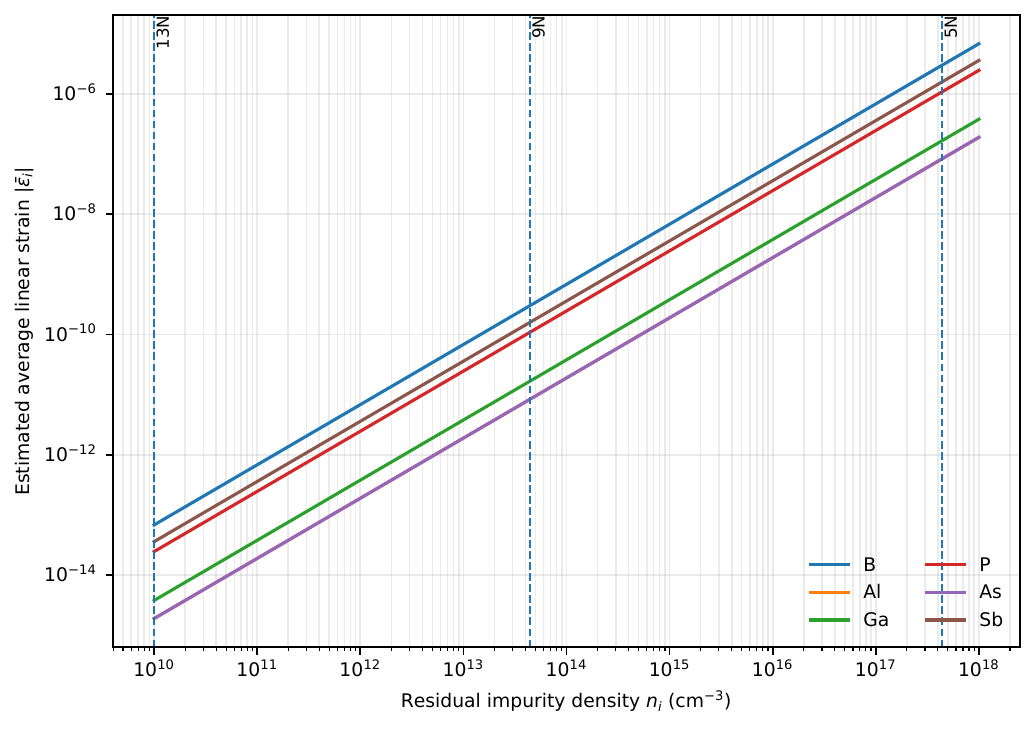}
    \caption{Estimated average linear strain in Ge as a function of residual impurity density for B, Al, Ga, P, As, and Sb, calculated from \( |\bar{\varepsilon}_i| \approx |\eta_i| n_i/N_0 \). The strain decreases linearly with impurity density, demonstrating how improved purity directly suppresses the background strain field. B, P, and Sb generate the largest strain because their size mismatch relative to Ge is largest. Only five distinct curves are visible because the Al and As curves overlap exactly: they have equal absolute mismatch parameters and therefore give the same \( |\bar{\varepsilon}_i| \). The dashed vertical lines mark representative impurity levels for 13N, 9N, and 5N Ge. The plot is intended as an order-of-magnitude comparative visualization rather than a precision materials prediction.}
    \label{fig:avg_strain_vs_impurity}
\end{figure}

This trend is visualized in Fig.~\ref{fig:avg_strain_vs_impurity}, which plots the estimated average linear strain as a function of residual impurity density for representative impurities using the scaling relation \( |\bar{\varepsilon}_i| \approx |\eta_i| n_i/N_0 \). Two important conclusions follow directly from the figure. First, the average strain decreases linearly with impurity density over many orders of magnitude, so improvements in chemical purity translate directly into lower background strain and reduced materials disorder. Second, the magnitude of the strain depends strongly on impurity species through the mismatch parameter \(\eta_i\): B and P generate substantially larger strain backgrounds than Al and Ga at the same impurity concentration because their size mismatch relative to Ge is larger. The vertical markers indicate representative impurity levels for detector-grade 13N Ge (\(\sim 10^{10}\,\mathrm{cm^{-3}}\)), 9N Ge (\(\sim 4.4\times10^{13}\,\mathrm{cm^{-3}}\)), and 5N Ge (\(\sim 4.4\times10^{17}\,\mathrm{cm^{-3}}\)). Figure~\ref{fig:avg_strain_vs_impurity} should therefore be interpreted as a comparative visualization, not as a precision prediction for a specific growth environment.

These materials considerations have direct implications for the qubit comparison developed in this paper. Donor and gate-defined electron qubits benefit immediately from reduced impurity disorder because it improves electrostatic uniformity, suppresses uncontrolled potential fluctuations, and makes valley-orbit physics easier to stabilize. Gate-defined hole qubits benefit because charge noise, interface disorder, and strain inhomogeneity all feed directly into the spin-orbit Hamiltonian and therefore into qubit-frequency variability. The strongest impact, however, is likely on acceptor qubits. Because acceptor-bound holes are exceptionally sensitive to local symmetry breaking and strain, their practical viability depends on a host crystal with extraordinarily small average and mesoscopic strain backgrounds. In this sense, the ability of Ge to be purified not only isotopically but also chemically---from commercial 5N feedstock through zone refining and ultimately to detector-grade single crystals with residual impurity densities near \(10^{10}\,\mathrm{cm^{-3}}\)---is not simply a materials advantage. It may be a key enabling factor that brings Ge acceptor qubits, as well as simpler bulk-Ge electron and hole qubit geometries, into more realistic reach \cite{Yang2015ZoneRefining,Wang2015USDHPGe,Bhattarai2024HPGeGrowth,AbadilloUriel2016}.

For completeness, Eq.~(\ref{eq:rmsstrain}) can also be used to estimate the strain background averaged over a qubit-sized region. For example, taking a characteristic averaging volume \(V\sim (100~\mathrm{nm})^3\), the worst-case random strain from B or Sb impurities in detector-grade 13N Ge is pushed down to the \(10^{-11}\) level, whereas 5N material remains near the \(10^{-7}\) range. Again, these values should be understood as order-of-magnitude indicators. Even so, the large separation between purity classes further illustrates why only very high-purity, low-dislocation Ge is a realistic host for the most strain-sensitive Ge qubit modalities.

\subsection{Phonon--charge and phonon--spin coupling in high-purity germanium}

The discussion in this subsection is intended as a compact comparative synthesis of coupling mechanisms relevant to the four Ge qubit classes, rather than as a full microscopic derivation for any single device geometry. High-purity Ge is relevant to all four qubit platforms not only because it suppresses chemical disorder and hyperfine broadening, but also because it enables long phonon propagation lengths and makes phononic band-structure engineering meaningful at the device scale \cite{Mei2025QST,Mei2025PDU}. For the present comparison, it is useful to distinguish two related but nonidentical couplings. The first is the direct phonon--charge coupling, more precisely a phonon--orbital or strain--orbital coupling, which determines how efficiently lattice motion perturbs localized charge states, tunnel barriers, valley splittings, or orbital energies. The second is the effective phonon--spin coupling, which arises only when a spin qubit inherits orbital character through spin--orbit interaction, valley mixing, or strain-dependent \(g\)-tensor modulation \cite{Li2012GeSpinLifetime,Terrazos2021,AbadilloUriel2023StrainSOI}. This distinction is important for device design: the phonon--charge channel is usually the faster transduction layer, while the phonon--spin channel is more selective and directly relevant to coherent control.

A compact way to summarize this hierarchy is
\begin{equation}
H_{\mathrm{ph\text{-}ch}}^{(e)} = H_{\mathrm{DP}}^{L}[\varepsilon(\mathbf{r},t)],
\qquad
H_{\mathrm{ph\text{-}ch}}^{(h)} = H_{\mathrm{BP}}[\varepsilon(\mathbf{r},t)],
\end{equation}
and
\begin{equation}
H_{\mathrm{ph\text{-}sp}}
\approx
\frac{\mu_B}{2}\,
\mathbf{B}\!\cdot\!
\left(\frac{\partial \mathbf{g}}{\partial \varepsilon}:\varepsilon_{\mathrm{ph}}\right)
\!\cdot\!
\bm{\sigma}
+
H_{\mathrm{mix}}[\varepsilon_{\mathrm{ph}}],
\end{equation}
where \(H_{\mathrm{DP}}^{L}\) denotes the deformation-potential coupling of L-valley electrons, \(H_{\mathrm{BP}}\) is the Bir--Pikus strain Hamiltonian acting on valence-band holes, and \(H_{\mathrm{mix}}\) collects strain-induced orbital and band-mixing terms that convert a nominally spin-like qubit into a spin-orbit-active one \cite{Li2012GeSpinLifetime,Terrazos2021,Wang2024Modeling}. In practice, \(H_{\mathrm{mix}}\) is relatively weak for electron-spin qubits and much stronger for hole-spin qubits, especially near heavy-hole/light-hole mixing points.

\paragraph{Direct phonon--charge coupling.}
For both electron- and hole-based Ge devices, the direct coupling of acoustic phonons to localized carriers is fundamentally stronger than the spin-selective coupling because it enters at first order through the strain tensor. For electrons this coupling primarily shifts L-valley orbital energies and modifies valley-orbit composition; for holes it acts directly on the valence-band manifold and therefore reshapes heavy-hole/light-hole mixing as well as the orbital spectrum \cite{Li2012GeSpinLifetime,Giorgioni2016GeQW,Terrazos2021,Wang2024Modeling}. This is one reason that hole-based devices are generally more strain-sensitive than electron-spin devices at comparable size.

The same direct phonon--charge channel is also the natural interface for phonon-based readout. In recent Ge phononic-spectroscopy and phonon-to-charge transducer concepts, guided acoustic excitations are first converted into a charge displacement or a conductance change in a localized state or RF-QPC-like sensor and are only then read out electrically \cite{Mei2025PDU}. For qubit design this means that phonons are most naturally suited to fast sensing and transduction through charge-like degrees of freedom, while fully spin-selective operations require an additional spin-orbit conversion step.

\paragraph{Electron-spin case.}
For electron spins in Ge, including donor spins and prospective gate-defined electron dots, the phonon--spin coupling is indirect. A phonon couples first to orbital, valley, or \(g\)-tensor degrees of freedom, and only then to the spin through spin--orbit admixture and valley repopulation \cite{Hasegawa1960,Roth1960,Li2012GeSpinLifetime}. This indirect structure has two consequences. First, the coherent phonon--spin coupling is generally weaker and narrower-band than in hole-spin platforms, so electron-spin devices are not the most natural candidates for very fast phonon-assisted control. Second, the same coupling can still dominate relaxation. That is already evident in donor spins in isotopically enriched Ge, where \(T_2\) becomes \(T_1\)-limited, showing that the spin-lattice channel is a central materials constraint rather than a small correction \cite{Sigillito2015}.

This electron-spin channel is therefore best characterized through relaxation rates, anisotropies, and cavity-mediated interaction scales rather than by a single universal vacuum-coupling number. In donor-based Ge, the large spin-orbit Stark shift demonstrates that the electron-spin Hamiltonian is strongly tunable by electric and strain-induced changes in the orbital environment \cite{Sigillito2016Stark}. In quantum-well conduction-electron systems, confinement can also strongly modify the electron \(g\) factor and spin lifetime \cite{Giorgioni2016GeQW}. Most importantly for phononic design, Smelyanskiy \emph{et al.} showed that donor spins in a Ge phononic crystal can exploit this same strong spin-lattice interaction in a controlled way: real one-phonon decay is suppressed inside a phononic bandgap, while virtual phonons still mediate purposeful long-range coupling. Their analysis emphasizes that the underlying Ge spin-lattice interaction is already much stronger than in bulk Si, which is precisely why bandgap engineering is so valuable for electron-spin platforms in Ge \cite{Smelyanskiy2014DonorGePhononic}.

\paragraph{Hole-spin case.}
For hole-spin qubits in Ge, including both acceptor-based and gate-defined hole devices, the phonon--spin coupling is qualitatively stronger because strain acts directly on a spin-orbit-active valence-band manifold. In practice, acoustic strain modulates heavy-hole/light-hole mixing, anisotropic \(g\) tensors, and the electric-dipole matrix elements responsible for EDSR-like control \cite{Terrazos2021,Wang2024Modeling,AbadilloUriel2023StrainSOI}. This is why Ge hole qubits support fast electrical manipulation and why they are particularly promising for coherent spin--phonon interfaces.

The strain sensitivity can be substantial even for modest gradients. Abadillo-Uriel \emph{et al.} showed that shear-strain gradients as small as \(3\times 10^{-6}\,\mathrm{nm}^{-1}\) can enhance hole-spin Rabi frequencies by about one order of magnitude, making clear that engineered strain should be treated as an active control resource rather than merely a fabrication imperfection \cite{AbadilloUriel2023StrainSOI}. Representative cavity-enhanced coupling scales are now also available. In the phononic-cavity architecture analyzed by Mei \emph{et al.}, Ge hole-spin qubits integrated with a phononic crystal cavity achieve predicted spin--phonon coupling strengths up to \(6.3~\mathrm{MHz}\), together with phononic-cavity quality factors above \(10^4\) and phonon-mediated \(T_1\) values reaching the millisecond scale when the acoustic density of states is engineered appropriately \cite{Mei2025QST}. These are not universal material constants, but they are realistic device-scale numbers that show why Ge holes are presently the strongest candidate for fast phonon-assisted control and phonon-bus coupling.

\paragraph{Implications for qubit control and readout.}
These material trends translate directly into platform-specific design logic. For electron-spin qubits in Ge, the safer strategy is usually to keep the qubit as spin-like as possible and use phonons either as a dissipation channel to suppress or as a carefully engineered virtual bus inside a phononic bandgap structure \cite{Sigillito2015,Smelyanskiy2014DonorGePhononic}. Fast measurement then still relies on spin-to-charge conversion followed by charge sensing, with phononic engineering playing a secondary but potentially powerful supporting role.

For hole-spin qubits, the design logic is more aggressive. Because the phonon--spin channel is already strong, phononic structures can be used proactively to enhance selected strain modes, suppress unwanted acoustic leakage, and tune the trade-off between fast control and long \(T_1\) \cite{Mei2025QST,AbadilloUriel2023StrainSOI}. In this sense, the broader systems lesson is simple: in Ge, phonon--charge coupling is the fast transduction layer for all four platforms, electron-spin phonon coupling is weaker and more naturally used selectively, while hole-spin phonon coupling is strong enough to become a primary control and coupling resource when the device is designed around strain gradients, sweet spots, and phononic bandgap engineering.

The next subsection turns this qualitative comparison into a common design framework for estimating how a PnC geometry modifies the longitudinal relaxation time \(T_1\) across Ge donor, acceptor, gate-defined electron, and gate-defined hole spin systems.

\subsection{Estimating \(T_1\) in PnC-engineered Ge spin systems}
\label{sec:pnc_T_1_general}

For all Ge-based spin-qubit modalities considered in this paper, phononic-crystal (PnC) engineering modifies relaxation by changing the acoustic modes available at the qubit transition frequency. The microscopic spin--phonon coupling mechanism is platform dependent. Donor and gate-defined electron spins couple to phonons primarily through strain-induced valley repopulation, orbital admixture, spin--orbit mixing, and \(g\)-tensor modulation, whereas acceptor and gate-defined hole spins couple more directly through the spin-\(3/2\) valence-band manifold, heavy-hole/light-hole mixing, and Bir--Pikus strain terms \cite{Roth1960,Hasegawa1960,AbadilloUriel2023StrainSOI,Mei2025QST}. Nevertheless, the design logic is common: a PnC can suppress the local strain density of states at the qubit transition frequency, thereby reducing the phonon-mediated contribution to \(1/T_1\), while surfaces, defects, microwave circuitry, and intentionally introduced cavity modes may impose new relaxation channels.

For a weakly coupled qubit with a well-defined transition frequency and a stationary phonon bath, the one-phonon relaxation rate can be written using the standard Fermi-golden-rule treatment. This starting point assumes that the qubit--bath coupling is perturbative and that a rate description is meaningful on the timescale of the relaxation measurement.

For a qubit transition of angular frequency
\begin{equation}
\omega_q = \frac{g_{\mathrm{eff}}\mu_B B_0}{\hbar},
\label{eq:omega_q_general}
\end{equation}
the phonon-mediated relaxation rate can be written schematically using Fermi's golden rule as
\begin{equation}
\Gamma_{1}^{\mathrm{ph}}
\equiv
\frac{1}{T_1^{\mathrm{ph}}}
=
\frac{2\pi}{\hbar}
\sum_{\lambda}
\left|
\left\langle 0;1_{\lambda}\right|
H_{\mathrm{q-ph}}
\left|1;0\right\rangle
\right|^2
\delta(\hbar\omega_q-\hbar\omega_{\lambda}),
\label{eq:golden_rule_t1_general}
\end{equation}
where \(|0\rangle\) and \(|1\rangle\) are the two qubit states, \(H_{\mathrm{q-ph}}\) is the effective qubit--phonon interaction, \(\lambda\) labels acoustic modes, and \(\omega_{\lambda}\) is the phonon frequency. Equation~\eqref{eq:golden_rule_t1_general} shows that the phonon-limited \(T_1\) is controlled by two factors: the platform-specific spin--phonon matrix element and the local acoustic density of states at the qubit transition frequency. In bulk Ge, the acoustic modes form a nearly continuous bath. In a PnC geometry, those modes can be strongly suppressed if \(\omega_q\) lies inside a phononic bandgap \cite{Smelyanskiy2014DonorGePhononic,Mei2025QST}.

For a compact cross-platform description, the one-phonon relaxation rate can be written in terms of the local strain cross-spectral density,
\begin{equation}
\Gamma_{1}^{\mathrm{ph}}(\omega_q)
=
\frac{1}{\hbar^2}
\sum_{\alpha\beta}
M_{\alpha}M_{\beta}^{*}
S_{\varepsilon_{\alpha}\varepsilon_{\beta}}(\mathbf r_q,\omega_q),
\label{eq:strain_spectral_rate}
\end{equation}
Here \(\mathbf r_q\) is the qubit position, \(\omega_q\) is the qubit transition angular frequency, and \(\alpha\) and \(\beta\) label strain-tensor components. The strain-coupling matrix element is
\[
M_{\alpha}
=
\left\langle 0\left|
\frac{\partial H_{\mathrm{q-ph}}}{\partial\varepsilon_{\alpha}}
\right|1\right\rangle .
\]
We use the unsymmetrized strain cross-spectral-density convention
\[
S_{\varepsilon_{\alpha}\varepsilon_{\beta}}(\mathbf r,\omega)
=
\int_{-\infty}^{\infty}
dt\,e^{i\omega t}
\left\langle
\delta\varepsilon_{\alpha}(\mathbf r,t)
\delta\varepsilon_{\beta}(\mathbf r,0)
\right\rangle ,
\]
where
\[
\delta\varepsilon_{\alpha}(\mathbf r,t)
\equiv
\varepsilon_{\alpha}(\mathbf r,t)
-
\left\langle\varepsilon_{\alpha}(\mathbf r,t)\right\rangle
\]
denotes the fluctuation of strain component \(\alpha\) about its mean value. This is the spectral-density convention used in the relaxation-rate expression above. Thus \(S_{\varepsilon_{\alpha}\varepsilon_{\beta}}\) describes the frequency-resolved correlation of the local strain fluctuations sampled by the qubit. The same convention is used for the patterned and reference structures.

We define the dimensionless, geometry- and polarization-weighted PnC suppression factor as
\begin{equation}
\eta_{\mathrm{PnC}}(\omega_q,\mathbf r_q)
=
\frac{
\sum_{\alpha\beta}
M_{\alpha}M_{\beta}^{*}
S^{\mathrm{PnC}}_{\varepsilon_{\alpha}\varepsilon_{\beta}}(\mathbf r_q,\omega_q)
}{
\sum_{\alpha\beta}
M_{\alpha}M_{\beta}^{*}
S^{\mathrm{ref}}_{\varepsilon_{\alpha}\varepsilon_{\beta}}(\mathbf r_q,\omega_q)
}.
\label{eq:spnc_general}
\end{equation}
Here \(S^{\mathrm{PnC}}_{\varepsilon_{\alpha}\varepsilon_{\beta}}\) is the local strain cross-spectral density in the patterned PnC device, while \(S^{\mathrm{ref}}_{\varepsilon_{\alpha}\varepsilon_{\beta}}\) is the corresponding quantity in a matched unpatterned reference/control structure. The reference structure has the same qubit modality, isotope composition, carrier or dopant density, surface treatment, device stack, magnetic-field orientation, and temperature as the PnC device. Therefore \(\eta_{\mathrm{PnC}}<1\) denotes suppression of the matrix-element-weighted local strain spectral density, \(\eta_{\mathrm{PnC}}=1\) denotes no change, and \(\eta_{\mathrm{PnC}}>1\) denotes enhancement. \(S^{\mathrm{ref}}_{\varepsilon_{\alpha}\varepsilon_{\beta}}\) is not assigned a universal numerical value; it is calculated or measured for the specified matched reference geometry.

If patterning does not appreciably change the qubit eigenstates or the matrix elements \(M_{\alpha}\), then
\begin{equation}
\Gamma_{1,\mathrm{PnC}}^{\mathrm{ph}}
\simeq
\eta_{\mathrm{PnC}}\Gamma_{1,\mathrm{ref}}^{\mathrm{ph}},
\qquad
T_{1,\mathrm{PnC}}^{\mathrm{ph}}
\simeq
\frac{T_{1,\mathrm{ref}}^{\mathrm{ph}}}{\eta_{\mathrm{PnC}}}.
\label{eq:pnc_rate_general}
\end{equation}
Here \(\Gamma_{1,\mathrm{ref}}^{\mathrm{ph}}=1/T_{1,\mathrm{ref}}^{\mathrm{ph}}\) is the one-phonon relaxation rate of the matched reference structure. As a donor-specific experimental example, the enriched Ge:P measurement with \(T_2=2T_1=1.2~\mathrm{ms}\) at \(B_0=0.44~\mathrm{T}\) and \(T=0.35~\mathrm{K}\) corresponds to \(T_1\simeq0.6~\mathrm{ms}\). This value is retained only as an illustrative Ge-donor reference point and is not transferred to acceptor, gate-defined hole, or gate-defined electron qubits.

As a design benchmark, \(\eta_{\mathrm{PnC}}\sim10^{-2}\) or \(10^{-3}\) would increase the phonon-limited \(T_1\) by roughly two or three orders of magnitude only when the suppressed one-phonon channel dominates.

A phononic crystal does not automatically lengthen \(T_1\). Its effect is determined by the matrix-element-weighted local strain spectral density at the qubit frequency. If the transition lies inside a phononic bandgap and one-phonon emission through the suppressed polarizations is dominant, \(T_1\) can increase. By contrast, a resonant defect mode can Purcell-enhance emission unless the qubit and mode are operated dispersively and the mechanical loss rate is sufficiently small. For donor and gate-defined electron spins, strain couples primarily through valley repopulation, \(g\)-tensor modulation, and spin--orbit-assisted orbital admixture. For acceptor and gate-defined hole spins, the dominant route is generally stronger Bir--Pikus coupling within the HH/LH manifold. Thus all four modalities can in principle benefit from phononic spectral engineering, but the attainable suppression and residual channels are platform and geometry dependent. Experiments in other solid-state systems have directly demonstrated phononic-crystal suppression of single-phonon relaxation and phononic-bandgap bath engineering, while an integrated Ge spin-qubit/PnC experiment remains to be demonstrated \cite{Kuruma2025,Odeh2025}.

The interpretation of \(H_{\mathrm{q-ph}}\) and \(\Gamma_{1,\mathrm{ref}}^{\mathrm{ph}}\) depends on the qubit modality. For Ge donor spins, the relevant low-temperature relaxation channel in bulk material is often the Roth--Hasegawa one-phonon spin-lattice process associated with spin--orbit coupling and the multivalley conduction-band structure of Ge \cite{Roth1960,Hasegawa1960}. In isotope-enriched Ge:P, pulsed ESR measurements have shown that donor-spin coherence can become relaxation-limited, with \(T_2=2T_1\) under the reported conditions \cite{Sigillito2015}. 
This donor-specific calibration is useful because it provides a concrete experimentally measured reference point. It should not, however, be transferred directly to acceptor or gate-defined hole qubits, where the spin--phonon matrix element depends strongly on gate bias, heavy-hole/light-hole mixing, confinement, strain, and the chosen operating point.
For gate-defined electron-spin qubits, the same general structure applies, but the spin--phonon matrix element is controlled by confinement, valley splitting, interface disorder, and spin--orbit admixture. For acceptor-spin and gate-defined hole-spin qubits, strain acts directly on the spin--orbit-active valence-band manifold, so \(H_{\mathrm{q-ph}}\) is often dominated by Bir--Pikus coupling, heavy-hole/light-hole mixing, and \(g\)-tensor modulation \cite{AbadilloUriel2023StrainSOI,Mei2025QST}. These differences affect the magnitude and anisotropy of \(T_1\), but not the basic role of the PnC suppression factor.

A practical estimate of \(T_1\) in a PnC-engineered Ge spin device should therefore proceed in three steps. First, the reference relaxation rate should be measured or calibrated in an unpatterned control sample with the same isotope composition, impurity or carrier density, surface treatment, device stack, magnetic-field orientation, and temperature range as the PnC device.

For shallow donor electrons in multivalley Ge, the direct one-phonon process can be described by the Roth--Hasegawa mechanism, in which acoustic strain changes the relative \(L\)-valley energies and, through spin--orbit-induced valley-dependent \(g\) tensors, modulates the donor-spin Hamiltonian \cite{Hasegawa1960,Roth1960}. In the high-temperature direct-process limit \(k_BT\gg\hbar\omega_q\) and for fixed field orientation, the resulting rate has the reduced scaling

\begin{equation}
\Gamma_{1}^{\mathrm{ph}}=\alpha(\theta)B_0^4T,
\label{eq:rh_scaling_general}
\end{equation}

where \(\alpha(\theta)\) contains the deformation-potential, elastic, valley, spin--orbit, and angular factors. Equation~\eqref{eq:rh_scaling_general} is the donor-specific Roth--Hasegawa direct-process scaling under the stated field, temperature, and valley-repopulation assumptions. It should not be applied unchanged to hole-spin devices or to regimes in which another relaxation mechanism dominates.

The reduced form assumes one-acoustic-phonon relaxation, a Zeeman energy small compared with valley--orbit excitation energies, and \(k_BT\gg\hbar\omega_q\), so that the Bose factor can be linearized. At lower temperature the full thermal occupation factor must be retained, while at higher temperature Raman or Orbach channels can dominate.

Second, the PnC suppression factor \(\eta_{\mathrm{PnC}}\) should be computed for the actual device geometry. This requires phononic band-structure and finite-structure calculations. The PnC unit cell is first simulated using Bloch-periodic boundary conditions to identify a complete or partial acoustic bandgap at the relevant qubit frequency. A finite PnC structure is then simulated to account for membrane thickness, the finite number of periods, fabrication disorder, clamping loss, and leakage through supports. The quantity needed for \(T_1\) is not only whether a bandgap exists, but the residual local strain density of states at the qubit location. A geometry-specific estimate is
\begin{equation}
\eta_{\mathrm{PnC}}(\omega_q,\mathbf{r}_q)
=
\frac{
\sum_{\lambda \in \mathrm{PnC}}
\left|
\mathcal{M}_{\lambda}^{(q)}(\mathbf{r}_q)
\right|^2
\delta_{\kappa}(\omega_q-\omega_{\lambda})
}{
\sum_{\lambda \in \mathrm{ref}}
\left|
\mathcal{M}_{\lambda,\mathrm{ref}}^{(q)}(\mathbf{r}_q)
\right|^2
\delta(\omega_q-\omega_{\lambda})
},
\label{eq:matrix_ldos_general}
\end{equation}
where \(\mathcal{M}_{\lambda}^{(q)}\) is the platform-specific strain-coupling matrix element for mode \(\lambda\), projected onto the qubit degree of freedom, and \(\delta_{\kappa}\) is a broadened delta function accounting for finite phonon linewidth, leakage, and mechanical damping. This expression is more general than a purely displacement-based density of states: it weights each phonon mode by the strain component and symmetry channel that actually couples to the qubit. For donor and electron-spin qubits, the relevant projections involve valley and \(g\)-tensor modulation; for acceptor and hole-spin qubits, they involve valence-band strain terms, heavy-hole/light-hole mixing, and electric-field-dependent spin--orbit coupling.

A shared localized or guided acoustic mode can also mediate a dispersive interaction
\begin{equation}
J_{ij}\simeq \frac{g_i g_j}{2}\left(\frac{1}{\Delta_i}+\frac{1}{\Delta_j}\right),
\qquad |\Delta_i|,|\Delta_j|\gg g_i,g_j,\kappa,
\label{eq:phonon_exchange_general}
\end{equation}
between qubits \(i\) and \(j\).

Here \(g_i\) (\(g_j\)) is the single-phonon coupling rate between qubit \(i\) (\(j\)) and the selected mechanical mode, \(\Delta_i=\omega_i-\omega_c\) and \(\Delta_j=\omega_j-\omega_c\) are the corresponding qubit--mode detunings, \(\omega_i\) and \(\omega_j\) are the qubit transition angular frequencies, and \(\omega_c\) is the mechanical-mode angular frequency. The quantity \(\kappa\) is the energy-decay linewidth of that mechanical mode, also expressed as an angular frequency. \(J_{ij}\) is the resulting coherent dispersive exchange rate between the two qubits. Equation~\eqref{eq:phonon_exchange_general} applies when \( |\Delta_i|,|\Delta_j|\gg g_i,g_j,\kappa\), so that the mode is only virtually populated.

For acceptor and gate-defined hole qubits, \(g_i\) arises mainly from strain-induced HH/LH mixing; for gate-defined electrons it is mediated through valley and orbital admixture. Device-specific calculations are presently most developed for donor and gate-defined hole concepts, and no phonon-mediated two-qubit gate has yet been demonstrated in an integrated Ge spin-qubit/PnC device.

Third, the total measured relaxation rate must include additional channels introduced by the device geometry. The experimentally measured \(T_1\) should be modeled as
\begin{equation}
\frac{1}{T_{1,\mathrm{meas}}}
=
\frac{
\eta_{\mathrm{PnC}}(\omega_q,\mathbf{r}_q)
}{
T_{1,\mathrm{ref}}^{\mathrm{ph}}
}
+
\Gamma_{\mathrm{surf}}
+
\Gamma_{\mathrm{def}}
+
\Gamma_{\mathrm{mw}}
+
\Gamma_{\mathrm{cav}}
+
\Gamma_{\mathrm{other}},
\label{eq:total_t1_general}
\end{equation}

Equation~\eqref{eq:total_t1_general} is a phenomenological additive-rate model that separates the suppressed one-phonon channel from parasitic channels introduced by surfaces, disorder, electrical noise, or fabrication. It is useful for design bookkeeping, but the individual parasitic rates remain device dependent.

Here \(\Gamma_{\mathrm{surf}}\) represents surface magnetic noise, charge noise, and surface two-level systems; \(\Gamma_{\mathrm{def}}\) represents implantation damage, etch damage, dislocations, and residual defects; \(\Gamma_{\mathrm{mw}}\) represents microwave, gate, or resonator-induced relaxation; and \(\Gamma_{\mathrm{cav}}\) represents relaxation through any intentionally or unintentionally introduced localized phonon mode. A PnC can increase \(T_1\) only until one of these non-bulk relaxation channels becomes dominant.

Special care is required if the PnC contains a defect cavity or guided mode near the qubit transition frequency. A phononic bandgap suppresses spontaneous emission into the acoustic continuum, but a resonant localized mode can enhance relaxation through a Purcell-like channel. If a cavity mode of frequency \(\omega_c\), linewidth \(\kappa=\omega_c/Q\), and qubit--phonon coupling \(g_{\mathrm{q-ph}}\) is present, the cavity-induced relaxation rate can be estimated as
\begin{equation}
\Gamma_{\mathrm{cav}}
\simeq
\frac{
g_{\mathrm{q-ph}}^2 \kappa
}{
\Delta^2+(\kappa/2)^2
},
\qquad
\Delta=\omega_q-\omega_c.
\label{eq:purcell_phonon_general}
\end{equation}
Therefore, a PnC designed primarily for long \(T_1\) should place the qubit transition inside the phononic bandgap and away from lossy localized modes. Conversely, if the device is designed to use a cavity or waveguide mode for coherent qubit--qubit coupling, then \(T_1\) protection and coherent coupling must be co-optimized.

A useful experimental strategy is to fabricate paired control and PnC-patterned samples with the same qubit modality, isotope composition, carrier or dopant density, surface treatment, and measurement conditions. The effective suppression factor can then be extracted directly from measured relaxation data:
\begin{equation}
\eta_{\mathrm{eff}}
\simeq
\frac{
\Gamma_{1,\mathrm{PnC}}-\Gamma_{\mathrm{other}}
}{
\Gamma_{1,\mathrm{control}}
}
\simeq
\frac{
T_{1,\mathrm{control}}
}{
T_{1,\mathrm{PnC}}
},
\label{eq:seff_exp_general}
\end{equation}
where the final approximation holds when non-phononic relaxation channels are small or unchanged between the two samples. For donor spins this comparison may be performed using pulsed ESR inversion-recovery measurements. For gate-defined electron and hole qubits, \(T_1\) is commonly measured by spin-to-charge conversion implemented through energy-selective tunnelling or Pauli-spin-blockade-based conversion, followed by charge detection with a nearby sensor dot or a dispersive RF gate-sensing circuit, depending on the device architecture. In all cases, the same structures should also be characterized by linewidth or \(T_2^*\), Hahn-echo \(T_2\), and dynamical-decoupling measurements.

This additional coherence characterization is necessary because increasing \(T_1\) does not guarantee a proportional increase in \(T_2\). The transverse coherence time obeys
\begin{equation}
\frac{1}{T_2}
=
\frac{1}{2T_1}
+
\frac{1}{T_{\phi}},
\label{eq:tphi_relation_general}
\end{equation}
where \(T_{\phi}\) is the pure-dephasing time. If a PnC extends \(T_1\) into the sub-second or seconds regime, then \(T_{\phi}\) may become the dominant coherence bottleneck. Thus, the central validation criterion for a PnC-protected Ge spin platform is not only an enhanced \(T_1\), but also a sufficiently long \(T_{\phi}\) after nanofabrication, membrane release, surface processing, and gate integration.

In summary, the PnC-modified relaxation time for any Ge-based spin system should be estimated by combining three ingredients: a calibrated reference relaxation rate for the relevant qubit modality, a geometry-specific local strain-density-of-states suppression factor, and additional relaxation channels introduced by surfaces, defects, cavities, and control circuitry. The working model is
\begin{equation}
T_{1,\mathrm{PnC}}
\simeq
\left[
\frac{
\eta_{\mathrm{PnC}}(\omega_q,\mathbf{r}_q)
}{
T_{1,\mathrm{ref}}^{\mathrm{ph}}
}
+
\Gamma_{\mathrm{other}}
\right]^{-1}.
\label{eq:working_model_general}
\end{equation}
This expression captures the common design principle across Ge donor, acceptor, gate-defined electron, and gate-defined hole platforms: a phononic bandgap can strongly increase the phonon-limited \(T_1\), but the experimentally realized relaxation time will be determined by the competition between phonon-density-of-states suppression and new relaxation channels introduced by nanofabrication and device integration.

\section{Donor spin qubits in Ge}

Donor-bound electron spins in Ge use shallow substitutional donors such as P, As, Sb, or Bi. Their atom-defined confinement, large electrical Stark response, and possible electron--nuclear registers make them complementary to lithographically defined Ge hole dots, while Ge's strong spin--orbit interaction and multivalley conduction band increase spin--phonon relaxation and make valley control important \cite{Sigillito2015,Sigillito2016Stark,Pica2016}.

\begin{figure}[htp!]
    \centering
    \includegraphics[width=0.98\linewidth]{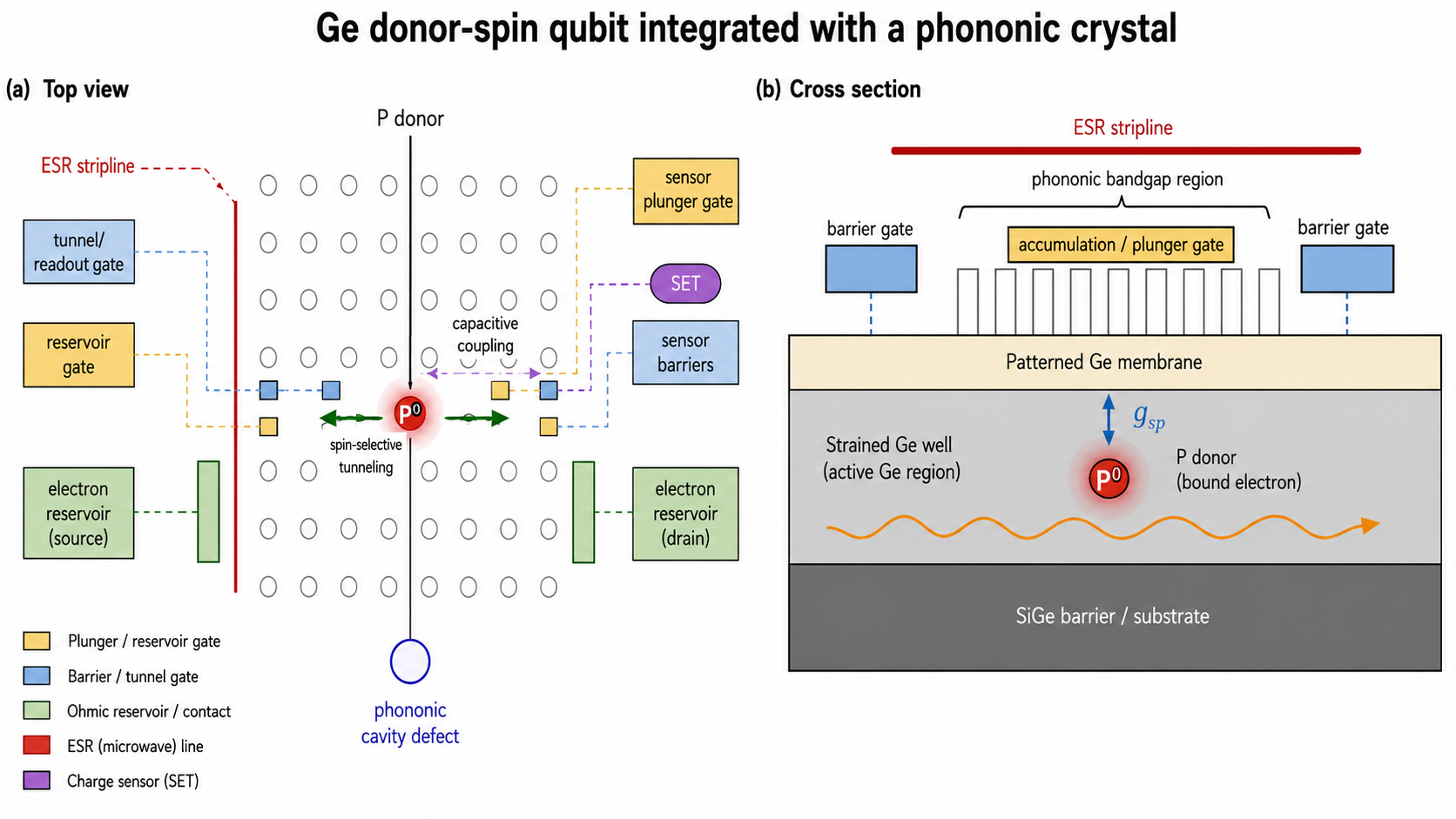}
    \caption{Conceptual Ge donor-spin device integrated with a phononic crystal. (a) A P donor is tunnel coupled to an electron reservoir through a readout/tunnel gate. A separate gate-defined sensor quantum dot or SET, with its own plunger and barrier gates and source/drain contacts, is capacitively coupled to the donor and detects spin-to-charge conversion. Direct ESR is driven by the labeled stripline, while a dc control gate Stark-shifts the donor resonance. Gray circles are etched phononic-crystal holes; the central modified-hole region is the phononic cavity defect and is not a metallic gate. (b) Cross-sectional schematic showing the patterned Ge membrane and donor location. The periodic PnC can place the donor transition in a phononic bandgap and suppress one-phonon emission. An optional high-$Q$ defect mode can provide a coherent interface or dispersive donor--donor coupling; if resonant and lossy, it can instead enhance relaxation. Orange waves denote elastic displacement or strain, and the blue double arrow denotes the local spin--phonon coupling $g_{sp}$.}
    \label{fig:ge_donor_pnc}
\end{figure}

Figure~\ref{fig:ge_donor_pnc} translates these physical ingredients into a device-level concept. The donor is tunnel coupled to an electron reservoir for energy-selective loading and readout, a separate sensor dot or SET detects spin-dependent charge transitions, and an on-chip stripline supplies the ac magnetic field for ESR. The patterned Ge membrane serves two spectrally distinct functions: the periodic region can suppress spontaneous one-phonon emission when the donor transition lies inside a bandgap, while an intentionally introduced defect or guided mode can provide a coherent spin--phonon interface when used in the dispersive regime.

Figure~\ref{fig:ge_donor_pnc} is an architecture-level schematic rather than a finalized fabrication layout. It separates the donor qubit, reservoir-assisted spin-to-charge conversion, independent charge sensing, ESR control, bandgap region, and optional localized mechanical mode so that the roles of the electronic and phononic subsystems are unambiguous.

\subsection{Physical encoding and projected Hamiltonian}

A schematic microscopic Hamiltonian for the donor electron is
\begin{equation}
H_{\mathrm{mic}}=\frac{1}{2}\mathbf p\!\cdot\!\mathbf m^{-1}\!\cdot\!\mathbf p+V_C(\mathbf r)+V_{\mathrm{cc}}^{(d)}(\mathbf r)+H_{\mathrm{SO}}+H_Z+H_{\mathrm{strain}}+e\mathbf E\!\cdot\!\mathbf r,
\label{eq:donor_microscopic}
\end{equation}

Here \(\mathbf r\) is the electron position measured from the donor site, \(\mathbf p=-i\hbar\nabla\) is the momentum operator, and \(\mathbf m^{-1}\) is the inverse anisotropic effective-mass tensor associated with the \(L\)-valley conduction band. \(V_C(\mathbf r)\) is the dielectric-screened donor Coulomb potential, and \(V_{\mathrm{cc}}^{(d)}(\mathbf r)\) is the short-range, donor-species-dependent central-cell correction that accounts for the breakdown of the smooth effective-mass description near the impurity core. \(H_{\mathrm{SO}}\) denotes the spin--orbit contribution, \(H_Z\) the Zeeman interaction with the applied magnetic field, and \(H_{\mathrm{strain}}\) the deformation-potential and valley-repopulation terms generated by strain. Finally, \(e>0\) is the elementary charge and \(\mathbf E\) is the applied electric field; the term \(e\mathbf E\!\cdot\!\mathbf r\) follows the electrostatic-potential convention used in Eq.~\eqref{eq:donor_microscopic}.

After solving or parameterizing the orbital problem and projecting onto a selected donor valley--orbit manifold, the effective qubit Hamiltonian may be written as
\begin{equation}
H_{\mathrm{donor}}^{\mathrm{eff}}=\mu_B\mathbf B\!\cdot\!\mathbf g\!\cdot\!\mathbf S+\mathbf I\!\cdot\!\mathbf A\!\cdot\!\mathbf S+H_{\mathrm{vo}}^{\mathrm{eff}}+H_{\mathrm{Stark}}^{\mathrm{eff}}+H_{\mathrm{strain}}^{\mathrm{eff}}+H_{\mathrm{sp}}^{\mathrm{eff}}.
\label{eq:donor_effective}
\end{equation}
The kinetic and binding energies are contained in the orbital eigenstates and their common energy offset, which is omitted from Eq.~\eqref{eq:donor_effective}. The listed terms are therefore effective operators within the projected donor manifold. In a realistic device, both the $g$ tensor and hyperfine tensor can depend on electric field, local strain, and valley repopulation \cite{Sigillito2016Stark,Pica2016}.

Several encodings are possible. The donor-bound electron spin provides the most direct qubit and supports magnetic-resonance control. A donor nuclear spin can provide a slower but potentially longer-lived memory, while the coupled electron--nuclear manifold can separate fast interface functions from long-lived storage within one impurity.

For a donor, the central-cell correction is a short-range modification of the screened Coulomb potential near the impurity core. It depends on donor species and determines the donor binding energy, valley--orbit splitting, and short-range wavefunction amplitude. Because the effective-mass approximation is not valid at atomic distances, this contribution must be calibrated to donor-specific spectroscopy or obtained from an atomistic treatment.

\subsection{Initialization, coherent control, and readout}

A donor electron can be initialized by energy-selective loading from a reservoir or by relaxation into the lower Zeeman state. Coherent rotations can be driven magnetically by electron-spin resonance or electrically when spin--orbit and Stark coupling provide an effective oscillating magnetic field. Readout converts the donor-spin state into a charge configuration that is detected by a nearby electrometer. Although demonstrated in silicon rather than Ge, the single-donor experiment of Pla \textit{et al.} provides a clear experimental template for this sequence of initialization, coherent control, and single-shot spin-to-charge readout \cite{Pla2012Donor}. The Ge implementation must additionally account for its stronger spin--orbit interaction and \(L\)-valley structure.

\subsection{Coherence, electrical tunability, and coupling}

As summarized in Sec.~2.1, the role of the dopant nuclear spin is set by isotope choice and by whether that degree of freedom is incorporated deliberately into the qubit architecture. Pulsed-ESR measurements established coherent donor spins in natural and isotopically enriched Ge. In samples containing appreciable $^{73}$Ge, spectral diffusion from the nuclear-spin bath limits $T_2$. In strongly enriched material, $T_2\simeq2T_1$, showing that spin--lattice relaxation becomes the coherence ceiling \cite{Sigillito2015}. Ge's stronger spin--orbit interaction enhances this phonon-assisted relaxation, but the same orbital admixture also produces exceptionally large and anisotropic donor spin Stark shifts. The donor resonance can be tuned by more than an ensemble linewidth, which is valuable for frequency allocation, selective addressing, and electrical placement of the Zeeman transition relative to a phononic bandgap or cavity mode \cite{Sigillito2016Stark}.

The conventional short-range entangling mechanism is exchange between neighboring donor-bound electrons. The larger and anisotropic donor envelopes in Ge can sustain useful exchange at larger separations than in Si, potentially relaxing placement tolerances, although multivalley interference, local strain, and electrostatic disorder remain important \cite{Pica2016}. In addition, two donors coupled dispersively to the same localized or guided phonon mode can acquire the interaction in Eq.~\eqref{eq:phonon_exchange_general}. This virtual-phonon route is a proposed architecture rather than a demonstrated Ge two-qubit gate \cite{Smelyanskiy2014DonorGePhononic}.

The dopant nuclear spin is a design-dependent resource or noise channel rather than an intrinsic platform advantage. A controllable nuclear-spin isotope can provide a long-lived electron--nuclear register. When the nuclear degree of freedom is not used, its influence can instead be mitigated through operating-point selection, dynamical decoupling, or ionization during protocol steps in which the bound electron is not required. Despite strong ensemble-level evidence, a single-donor Ge device with a complete set of initialization, single-shot readout, one-qubit fidelity, and two-qubit-gate benchmarks has not yet been demonstrated.

\section{Acceptor spin qubits in germanium}

An acceptor in Ge binds a valence-band hole and therefore inherits the spin-$3/2$ HH/LH manifold, strong spin--orbit coupling, quadrupolar response, and pronounced sensitivity to strain and interface symmetry. This creates an impurity-defined qubit with unusually strong electrical and mechanical tunability, but also stringent requirements for microscopic reproducibility \cite{AbadilloUriel2016,Salfi2016QuantumComputing,Salfi2016ChargeInsensitive,AbadilloUriel2023StrainSOI}.

\subsection{Physical encoding and low-energy Hamiltonian}

In bulk Ge, the acceptor spectrum is set by the Kohn--Luttinger valence-band structure dressed by the impurity central-cell potential. Near a device interface, broken inversion symmetry, dielectric mismatch, vertical confinement, local strain, and gate electric fields split the nominally fourfold ground manifold into two Kramers doublets with electrically tunable HH/LH admixture \cite{AbadilloUriel2016}. One doublet can be selected as the computational basis, while the nearby excited doublet enables electrical driving, spin--orbit mixing, and coupling to microwave or strain fields.

A schematic Hamiltonian is
\begin{equation}
H_{\mathrm{acc}}=H_{\mathrm{KL}}+H_{\mathrm{cc}}+H_{\mathrm{int}}+H_{\mathrm{BP}}+H_E+H_B,
\label{eq:acceptor_hamiltonian}
\end{equation}
where $H_{\mathrm{KL}}$ is the Kohn--Luttinger Hamiltonian, $H_{\mathrm{cc}}$ is the central-cell contribution, $H_{\mathrm{int}}$ represents interface confinement and symmetry breaking, $H_{\mathrm{BP}}$ is the Bir--Pikus strain term, and $H_E$ and $H_B$ describe electric- and magnetic-field coupling. The qubit eigenstates are therefore determined not only by the impurity species, but by the complete local electrostatic and elastic environment \cite{AbadilloUriel2016,Zhang2023AcceptorStrain}.

For an acceptor, central-cell physics acts on a bound valence-band hole rather than an \(L\)-valley conduction electron. The short-range impurity potential therefore modifies the composition and splittings of the \(J=3/2\) acceptor manifold and can change its response to strain, interfaces, and electric fields. Its form is acceptor-species dependent and is not interchangeable with the donor central-cell correction. We consequently include the acceptor central-cell term only in the acceptor Hamiltonian and discuss it together with the symmetry-breaking terms that split and mix the acceptor Kramers doublets.

\begin{figure}[htp!]
    \centering
    \includegraphics[width=0.98\linewidth]{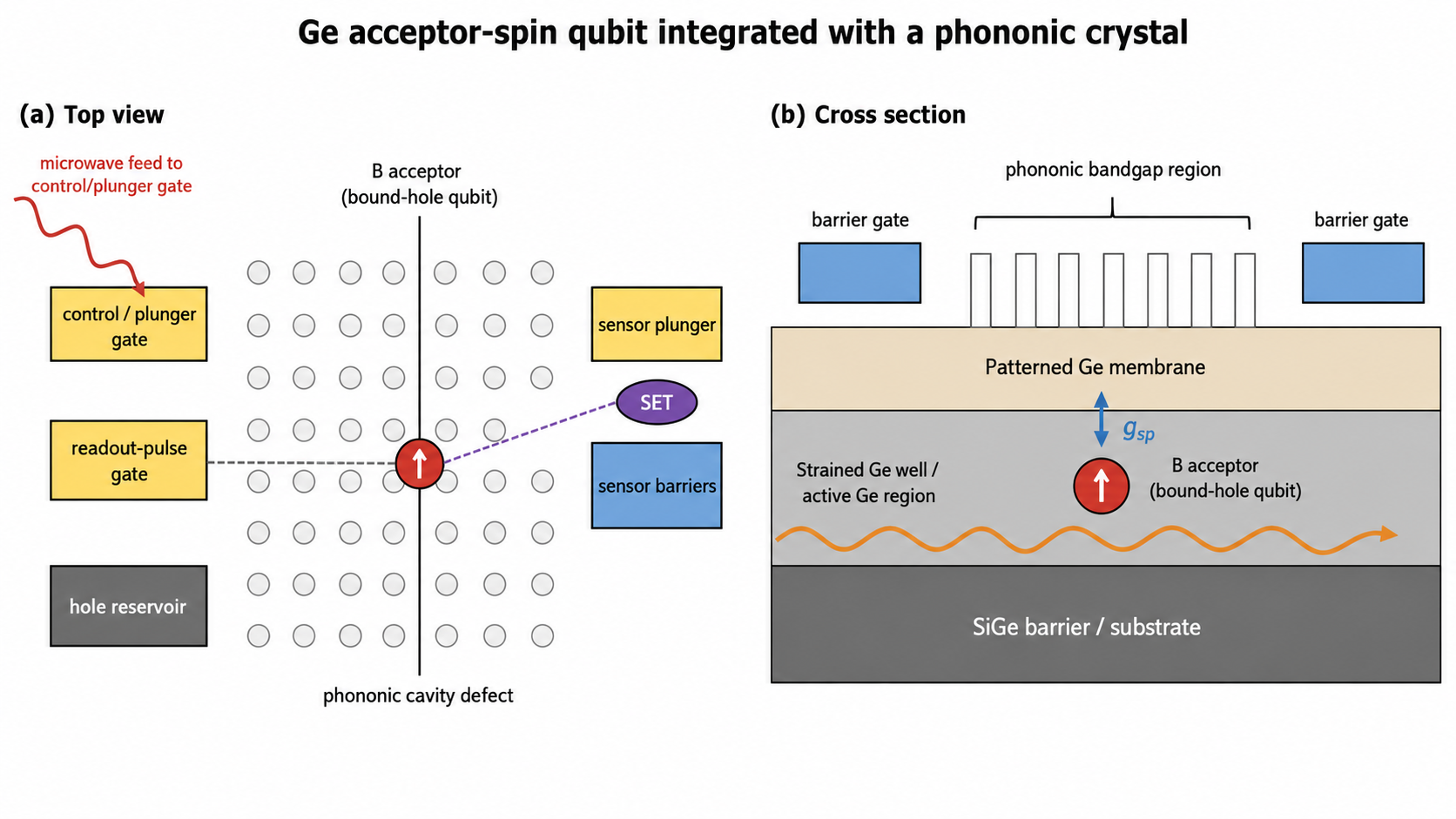}
    \caption{Conceptual Ge acceptor-spin device integrated with a phononic crystal. (a) A near-interface B acceptor is coupled to a hole reservoir and controlled by a dc control/plunger gate with a microwave feed for EDSR. A separate readout-pulse gate makes acceptor--reservoir tunnelling spin selective, and a gate-defined sensor dot or SET detects the charge transition. The red acceptor symbol with an internal arrow denotes one B acceptor hosting the bound-hole pseudospin. Gray circles are phononic-crystal holes, and the central modified-hole region is the cavity defect. (b) Geometric cross section through the gate stack, patterned Ge membrane, and near-interface acceptor. The acceptor symbol marks the spatial position of the bound-hole state, while the shaded and wavelike features denote the localized strain field of the phononic structure.}
    \label{fig:GeAcceptorSpinPlatform}
\end{figure}

Figure~\ref{fig:GeAcceptorSpinPlatform} shows a near-interface acceptor device in which the control, readout, sensor, and phononic functions are physically separated. The figure also makes clear that the acceptor symbol and its internal pseudospin arrow denote one bound-hole qubit, not two different acceptors.

\subsection{Initialization, coherent control, readout, and coupling}

A dc gate field selects and tunes the low-energy Kramers doublet, and an ac voltage applied to the control/plunger gate in Fig.~\ref{fig:GeAcceptorSpinPlatform} can drive EDSR through HH/LH and quadrupolar mixing. Initialization and readout may be implemented by energy-selective tunnelling between the acceptor and a nearby hole reservoir, with the resulting charge transition detected by the sensor dot or SET. Candidate two-qubit interactions include short-range exchange, electric-dipole-mediated coupling, and dispersive coupling through a shared strain mode. For two acceptors coupled to one mechanical mode, the effective interaction scales as $g_1g_2/\Delta$, with $g$ arising primarily from strain-induced HH/LH mixing. These operations remain prospective for a complete Ge acceptor-qubit stack.

\subsection{Physical opportunities and materials challenges}

Acceptor qubits combine chemically defined localization with a spin--orbit-active valence-band state. Interface-induced parity mixing can enable EDSR, electrically tunable selection rules, charge-insensitive operating points, and switchable electric dipoles \cite{Salfi2016QuantumComputing,Salfi2016ChargeInsensitive,AbadilloUriel2018MagicAngles}. Their spin-$3/2$ character also provides quadrupolar couplings, anisotropic $g$ tensors, and strong coupling to strain, making acceptors attractive for phononic, microwave, and other hybrid interfaces \cite{AbadilloUriel2023StrainSOI}.

As summarized in Sec.~2.1, the role of the dopant nuclear spin is set by isotope choice and by whether that degree of freedom is incorporated deliberately into the qubit architecture. The same physics makes the platform sensitive to central-cell chemistry, dopant depth, dielectric mismatch, interface fields, and local strain. These are not small perturbations; they reshape the qubit manifold and can create substantial device-to-device variability \cite{AbadilloUriel2016,Zhang2023AcceptorStrain}. Strong electrical and strain response must therefore be combined with controlled impurity placement, high-quality interfaces, and operation near favorable sweet spots. A suitable nuclear-spin-zero acceptor species or isotope, if compatible with the desired Ge device architecture, would remove the local dopant hyperfine channel; otherwise an uncontrolled dopant nucleus can add dephasing, while a deliberately addressable nucleus could instead become a hybrid-register resource.

\subsection{Experimental evidence and present status}

The most developed acceptor-related work remains in Si-compatible platforms. Salfi \textit{et al.} introduced theoretical proposals for near-interface single-acceptor spin--orbit qubits and electrically tunable sweet spots \cite{Salfi2016QuantumComputing,Salfi2016ChargeInsensitive}. Experimental studies instead established manipulation and decoherence of acceptor states in Si and resolved the spin states and relaxation of a single-acceptor transistor \cite{Song2011,vanDerHeijden2014}. These experiments support the underlying acceptor-spin physics, but a fully controlled Ge acceptor qubit with benchmarked coherent gates has not yet been demonstrated. The architecture in Fig.~\ref{fig:GeAcceptorSpinPlatform} should therefore be interpreted as a testable Ge device concept rather than an established processor technology.

\section{Gate-defined hole spin qubits in strained Ge/SiGe heterostructures}

Gate-defined Ge hole qubits use electrostatic confinement in Ge nanowires or, increasingly, strained planar Ge/SiGe quantum wells. The logical doublet derives from the valley-free $\Gamma$-point valence band. Confinement, compressive strain, and gate fields select a predominantly HH-like state with controlled HH/LH mixing, enabling strong EDSR and electrically tunable $g$ tensors \cite{Scappucci2021,Terrazos2021,Wang2021Optimal}.

\begin{figure}[htp!]
    \centering
    \includegraphics[width=0.98\linewidth]{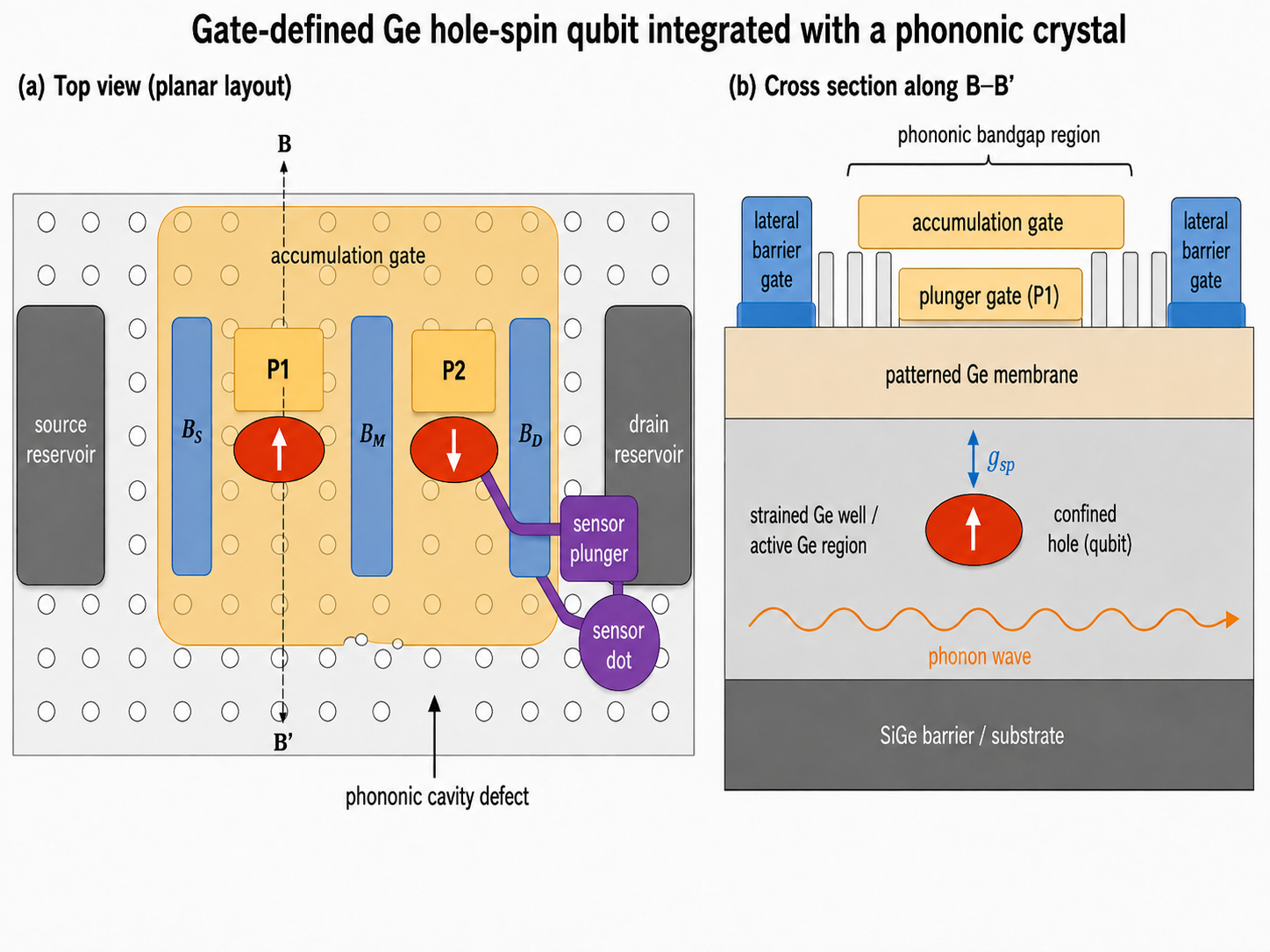}
    \caption{Conceptual gate-defined Ge hole-spin device integrated with a phononic crystal. (a) Plan view of two neighboring gate-defined Ge hole-spin qubits. The semitransparent accumulation gate spans the active channel. The B--B$^{\prime}$ section line passes through the center of the P1 plunger gate and the corresponding confined-carrier density, so the accumulation gate, P1 plunger gate, and confined carrier shown in panel (b) correspond directly to their projected positions in panel (a); reduced opacity is used only to keep the local gates and confined-carrier density visible. (b) Cross section through the left qubit along B--B$^{\prime}$, showing the vertical metal/dielectric/Ge--SiGe stack and the confined hole directly beneath its plunger gate. The second qubit in panel (a) is outside the section plane. Gray circles are phononic-crystal holes, the modified central region is the cavity defect, orange waves denote elastic displacement or strain, and the blue double arrow denotes the local spin--phonon coupling $g_{sp}$.}
    \label{fig:GeGateDefinedHoleSpinPlatform}
\end{figure}

Figure~\ref{fig:GeGateDefinedHoleSpinPlatform} shows the overlapping gate architecture required to accumulate a hole channel and define quantum dots. The metal electrodes are labeled by their electrical function. An accumulation gate attracts carriers to the Ge channel and establishes the conducting layer or reservoir. A plunger gate primarily tunes the electrochemical potential and occupation of an individual quantum dot and may also carry the oscillating electric field used for EDSR. A barrier gate controls the tunnel rate between a dot and a reservoir or between neighboring dots and therefore tunes exchange coupling. Gates associated with a nearby sensor dot tune its conductance or RF response for charge detection. The red ellipses in the schematic denote the spatial probability density of the confined carrier states; they are not additional electrodes.

\subsection{Physical encoding and device operation}

A schematic effective Hamiltonian is
\begin{equation}
H_{\mathrm{hQD}}=H_{\mathrm{LK}}+H_{\mathrm{BP}}+V_{\mathrm{conf}}(\mathbf r,\mathbf E)+\frac{1}{2}\mu_B\mathbf B\!\cdot\!\mathbf g(\mathbf E)\!\cdot\!\bm\sigma+H_{\mathrm{ac}}(t)+H_{\mathrm{q-ph}},
\label{eq:hole_qd_hamiltonian}
\end{equation}
where the Luttinger--Kohn and Bir--Pikus terms describe the valence manifold and strain, while the gate-dependent confinement controls the orbital state, HH/LH admixture, and qubit $g$ tensor.

The accumulation/plunger/barrier stack shown in Fig.~\ref{fig:GeGateDefinedHoleSpinPlatform} first induces a two-dimensional hole gas and then isolates a controlled odd occupancy in each quantum dot. A microwave voltage applied to a plunger gate produces EDSR through the electrically tunable spin--orbit interaction.

Initialization and readout can use spin-to-charge conversion through two common routes. In energy-selective tunnelling, the spin-dependent alignment of a dot level with a reservoir converts the spin state into a transient charge event, following the single-spin readout principle introduced by Elzerman \textit{et al.} \cite{Elzerman2004Readout}. In a double dot, Pauli spin blockade or parity conversion maps the two-spin state onto an allowed or blocked charge transition; the relevant conditions in spin--orbit-coupled silicon devices are analyzed by Seedhouse \textit{et al.} \cite{Seedhouse2021PauliBlockade}. In either case, the resulting charge configuration is detected by a nearby sensor dot or by RF reflectometry.

Two-qubit gates are implemented by pulsing the interdot barrier or detuning to control exchange, preferably near a charge-symmetry point. A shared localized or guided acoustic mode can additionally generate a dispersive interaction through strain-induced HH/LH mixing, but such a phonon-mediated gate has not yet been demonstrated in an integrated Ge device.

\subsection{Control advantages and engineering challenges}

The valley-free valence-band edge, light in-plane mass, and strong spin--orbit interaction make gate-defined Ge holes especially favorable for dense, all-electrical quantum-dot circuits. The same heterostructure can support single dots, double dots, charge sensors, exchange-coupled qubit pairs, and larger arrays. Theory and experiment also show that strong EDSR can coexist with reduced first-order noise sensitivity at suitable operating points \cite{Wang2021Optimal,Hendrickx2024Sweet}.

Fast control is not free. Charge noise, dielectric defects, gate-voltage drift, interface disorder, and unintended variations in confinement can shift the qubit frequency and Rabi rate. The states are HH dominated but not purely HH, so the $g$ tensor, driving strength, and relaxation rate depend strongly on electric field, dot geometry, strain, and magnetic-field orientation \cite{Wang2021Optimal,Hendrickx2024Sweet}. Integrating the PnC shown in Fig.~\ref{fig:GeGateDefinedHoleSpinPlatform} adds a further co-design constraint: the patterned membrane must engineer the acoustic spectrum without degrading electrostatic confinement, interface quality, or charge stability.

Gate-defined holes are the most experimentally mature Ge modality. The quantitative benchmarks, operating temperatures, and demonstrated array scales are consolidated in Table~\ref{tab:quantitative_comparison} to avoid repeating a separate maturity assessment in this platform section.

\section{Gate-defined electron spin qubits in germanium}

\subsection{Motivation and low-energy Hamiltonian}

Gate-defined Ge electron qubits offer post-fabrication electrostatic control of the qubit position, occupancy, tunnel coupling, and exchange interaction. A dot can be emptied and reloaded, coupled controllably to reservoirs and charge sensors, shuttled through an array, and incorporated into one- or two-dimensional layouts using the same gate stack. These capabilities distinguish electrostatic dots from impurity-bound donor devices, for which the atomic position and species are fixed during fabrication. The resulting architectural flexibility must be weighed against stronger sensitivity to interfaces, charge noise, and the multivalley \(L\)-point spectrum.

Strain and band-offset engineering in Ge/SiGe can create a conduction-electron system that is not simply bulk Ge in miniature. Experimental work on Ge/SiGe quantum wells has demonstrated confinement-induced engineering of the $L$-valley electron $g$ factor and spin lifetime, providing a materials basis for a source-loaded electron channel \cite{Giorgioni2016GeQW}. The central difficulty is that valley degeneracy, anisotropic masses, intervalley coupling, and strain dependence remain leading components of the device physics.

A schematic single-electron Hamiltonian is
\begin{equation}
H_{\mathrm{eQD}}=H_{\mathrm{conf}}+H_{\mathrm{orb}}+H_{\mathrm{valley}}+\frac{1}{2}\mu_B\mathbf B\!\cdot\!\mathbf g\!\cdot\!\bm\sigma+H_{\mathrm{SO}}(E)+H_{\mathrm{ac}}(t)+H_{\mathrm{q-ph}}.
\label{eq:electron_qd_hamiltonian}
\end{equation}
Gate-defined Ge hole qubits derive from the $\Gamma$-point valence band and therefore do not carry the $L$-valley degree of freedom relevant to conduction electrons. In an electron dot, valley energies, intervalley mixing, and valley-dependent masses are leading components of the low-energy orbital Hamiltonian rather than perturbative corrections \cite{Baron2003Lvalleyg,Giorgioni2016GeQW,Virgilio2009GeValley}.

\begin{figure}[htp!]
    \centering
    \includegraphics[width=0.98\linewidth]{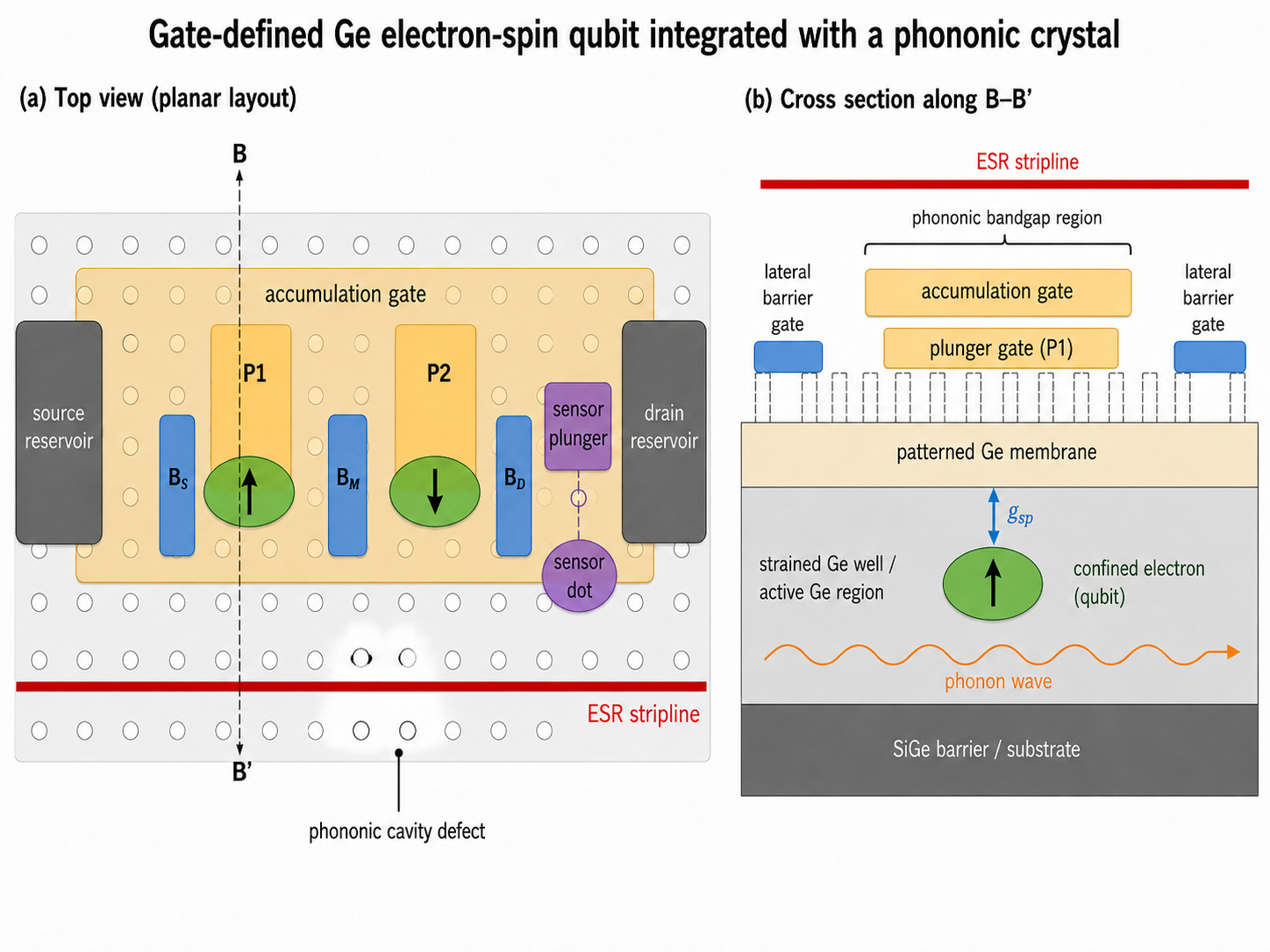}
    \caption{Conceptual gate-defined Ge electron-spin device integrated with a phononic crystal. (a) Plan view of two neighboring gate-defined Ge electron-spin qubits using the same drawing convention as Fig.~\ref{fig:GeGateDefinedHoleSpinPlatform}. 
    The semitransparent accumulation gate spans the active channel. The B--B$^{\prime}$ section line passes through the center of the P1 plunger gate and the corresponding confined-carrier density, so the accumulation gate, P1 plunger gate, and confined carrier shown in panel (b) correspond directly to their projected positions in panel (a); reduced opacity is used only to keep the local gates and confined-carrier density visible. (b) Cross section through the left qubit, showing the vertical gate stack and the confined electron beneath its plunger gate; the second dot is outside the section plane. The baseline control path is a separate ESR stripline; electrically driven alternatives depend on the realized spin--orbit and valley-mixing regime. Gray circles are phononic-crystal holes, the modified central region is the cavity defect, orange waves denote elastic displacement or strain, and the blue double arrow denotes the local spin--phonon coupling $g_{sp}$.}
    \label{fig:GeGateDefinedElectronSpinPlatform}
\end{figure}

Figure~\ref{fig:GeGateDefinedElectronSpinPlatform} shows a multilayer gate stack that accumulates and confines electrons, independently defines a sensor dot, and provides a direct-ESR control path. The metal electrodes are labeled by their electrical function. An accumulation gate attracts electrons to the Ge channel and establishes the conducting layer or reservoir. A plunger gate primarily tunes the electrochemical potential and occupation of an individual quantum dot. A barrier gate controls the tunnel rate between a dot and a reservoir or between neighboring dots and therefore tunes exchange coupling. Gates associated with a nearby sensor dot tune its conductance or RF response for charge detection. The green ellipses denote the spatial probability density of the confined electron states; they are not additional electrodes.

\subsection{Initialization, control, coupling, and experimental outlook}

The gate stack in Fig.~\ref{fig:GeGateDefinedElectronSpinPlatform} accumulates electrons and defines few-electron dots. Spin rotations could be driven by the ESR stripline, or electrically through $g$-tensor modulation or spin--orbit/valley-assisted EDSR if a suitable regime is established. Initialization and readout would use spin-selective tunnelling or Pauli spin blockade with the proximal sensor dot, and two-qubit gates would most naturally use electrically tunable exchange. Coupling through a shared phonon mode is possible in principle through valley and orbital admixture, but is more indirect than for hole-like states and remains prospective.

Gate-defined electrons and holes exchange one form of internal complexity for another. An electron provides a nominal spin-$1/2$ logical basis only after strain, confinement, and interfaces produce a stable valley splitting that isolates one $L$-valley-derived orbital state. Otherwise, valley degeneracy, intervalley mixing, anisotropic masses, and valley-dependent $g$ tensors introduce leakage and variability. Hole qubits avoid $L$ valleys but retain HH/LH and spin--orbit manifold complexity. Neither platform is generically simpler without specifying the operating regime.

The measured confinement-induced $g$-factor engineering of $L$-valley electrons suggests that electrical tuning of the Zeeman response is realistic, but a mature gate-defined Ge electron-qubit stack with established initialization, coherent control, readout, relaxation, and fidelity benchmarks has not yet emerged \cite{Baron2003Lvalleyg,Giorgioni2016GeQW}. Figure~\ref{fig:GeGateDefinedElectronSpinPlatform} should therefore be read as a credible experimental design direction rather than a representation of an already benchmarked processor.

\section{Cross-platform comparison}

To reduce repetition, Sections 3--6 emphasize physical encoding, device implementation, initialization, coherent control, readout, and platform-specific coupling mechanisms. Comparative advantages, limitations, experimental maturity, and quantitative benchmarks are consolidated in this section. This organization avoids repeating the same assessment within every platform section and enables direct comparison on a common basis.

\subsection{A physics-based comparison}
Table~\ref{tab:comparison} summarizes the main trade-offs among the four principal Ge-based qubit modalities discussed above. This comparison should be read with two caveats. First, no modality is universally ``best'' independent of application: a qubit optimized for long-lived quantum memory, a dense processor tile, and a phonon-coupled hybrid transducer need not be the same physical object. Second, the four platforms are not equally mature experimentally. Gate-defined hole qubits in strained Ge/SiGe have already reached a genuinely multi-qubit and processor-relevant regime, whereas donor, acceptor, and gate-defined electron qubits in Ge remain at earlier stages of development. Accordingly, the comparison below is intended to be physics-based rather than purely performance-based. It weighs what each modality appears to offer in principle while also accounting for what has and has not yet been demonstrated experimentally. Here, the qualitative maturity labels refer to the current level of experimental development, including demonstrated single-qubit control, two-qubit operations where available, multi-qubit integration, and broader architecture-level readiness.

\begin{table}[H]
\caption{Technical comparison of four Ge-based qubit modalities. ``Maturity'' is a qualitative assessment as of 2026. Isotope-dependent dopant hyperfine coupling is treated as a controllable resource or noise channel rather than as an intrinsic platform advantage or disadvantage.}
\label{tab:comparison}
\centering
\footnotesize
\setlength{\tabcolsep}{4pt}
\renewcommand{\arraystretch}{1.18}

\begin{tabularx}{\linewidth}{
>{\raggedright\arraybackslash}p{0.18\linewidth}
>{\raggedright\arraybackslash}X
>{\raggedright\arraybackslash}X
>{\raggedright\arraybackslash}X
>{\raggedright\arraybackslash}p{0.13\linewidth}
}
\toprule
\textbf{Modality} &
\textbf{Best features} &
\textbf{Main liabilities} &
\textbf{Control paradigm} &
\textbf{Maturity} \\
\midrule

Donor spin qubit &
Atom-defined confinement; large Stark tunability; possible electron--nuclear memory &
Spin--lattice relaxation in Ge; L-valley complexity; deterministic placement challenge &
ESR/EDSR with hyperfine and Stark tuning &
Early--intermediate \\

Acceptor spin qubit &
Atom-like spin-$3/2$ hole; strong electric and strain response; phonon/photon coupling potential &
Central-cell, strain, and interface sensitivity; isotope-dependent local hyperfine channel; limited Ge demonstrations &
Electric control of Kramers doublets; strain/interface tuning &
Early \\

Gate-defined hole-spin qubit &
All-electric control; valley-free valence band; multiqubit progress; scalable arrays &
Charge-noise sensitivity; $g$-tensor anisotropy; materials and interface variability &
EDSR, exchange gates, singlet--triplet control, and shuttling &
Advanced \\

Gate-defined electron spin qubit &
Post-fabrication reconfigurability of dot position, occupancy, tunnel coupling, and exchange; compatibility with Si-style gate concepts &
L-valley complexity; limited Ge-specific maturity; unclear advantage over holes &
Baseline ESR; possible $g$-tensor- or valley-assisted electrical control; exchange gates &
Early \\

\bottomrule
\end{tabularx}
\end{table}
\FloatBarrier

\subsection{Quantitative benchmarks and silicon context}

Table~\ref{tab:quantitative_comparison} compiles representative experimental benchmarks for the four Ge modalities and selected Si electron- and hole-spin results. Values obtained with different magnetic fields, temperatures, devices, pulse definitions, and readout protocols are not directly equivalent material constants; missing entries identify measurements that have not yet been demonstrated rather than zero performance.

\begingroup
\footnotesize
\setlength{\tabcolsep}{4pt}
\renewcommand{\arraystretch}{1.16}

\begin{xltabular}{\textwidth}{
    @{}
    >{\raggedright\arraybackslash}p{0.20\textwidth}
    Y
    >{\raggedright\arraybackslash}p{0.23\textwidth}
    @{}
}
\caption{Representative quantitative benchmarks for Ge spin-qubit
modalities, with selected Si electron- and hole-spin results for context.
``DD'' denotes dynamical decoupling. Values are representative
experimental results obtained with different magnetic fields,
temperatures, devices, pulse definitions, and readout protocols;
they should not be interpreted as directly equivalent material
constants. Missing benchmarks indicate that the corresponding
coherent Ge qubit demonstration has not yet been reported.}
\label{tab:quantitative_comparison}\\

\toprule
Benchmark category &
Representative result &
Status and interpretation \\
\midrule
\endfirsthead

\multicolumn{3}{c}{
    \textit{Table~\thetable\ continued}
}\\
\toprule
Benchmark category &
Representative result &
Status and interpretation \\
\midrule
\endhead

\midrule
\multicolumn{3}{r}{
    \textit{Continued on next page}
}\\
\endfoot

\bottomrule
\endlastfoot

\multicolumn{3}{@{}l}{
    \textbf{Ge donor electrons---ensemble ESR demonstrated}
}\\
\addlinespace[0.3em]

One-qubit control &
No single-donor gate fidelity has yet been reported in Ge.
Electrical Stark tuning exceeds the ensemble linewidth
\cite{Sigillito2016Stark}. &
Strong materials-level evidence exists, but single-donor
initialization, coherent control, and single-shot readout remain
to be demonstrated. \\

Two-qubit control &
No Ge donor two-qubit gate has been demonstrated. Electrically
controlled exchange and phonon-mediated interactions remain
proposed approaches
\cite{Pica2016,Smelyanskiy2014DonorGePhononic}. &
No fabricated donor-based multiqubit Ge array has yet been
reported. \\

Coherence and relaxation &
Strongly \(^{73}\)Ge-depleted ensembles exhibit
\(T_2^*=0.21\)--\(0.28~\mu\mathrm{s}\). The longest reported
Hahn-echo value is \(T_2=1.2~\mathrm{ms}\), with
\(T_2=2T_1\), implying \(T_1\simeq0.6~\mathrm{ms}\), at
\(0.35~\mathrm{K}\) and \(0.44~\mathrm{T}\)
\cite{Sigillito2015}. &
The reported values were obtained from bulk donor ensembles,
not individually controlled donor qubits. \\

\midrule

\multicolumn{3}{@{}l}{
    \textbf{Ge acceptor-bound holes---pre-benchmark platform}
}\\
\addlinespace[0.3em]

One-qubit control &
No coherently benchmarked Ge acceptor qubit has yet been
reported. &
Electrical control is theoretically attractive because of
spin-\(3/2\), heavy-hole/light-hole, and quadrupolar mixing. \\

Two-qubit control &
No two-qubit acceptor gate has been demonstrated in Ge. &
Exchange, electric-dipole coupling, and shared strain-mode
coupling remain proposed mechanisms. \\

Coherence and relaxation &
A complete experimental set of \(T_1\), \(T_2^*\), echo
\(T_2\), gate times, and gate fidelities is not yet available
for a Ge acceptor qubit. Experiments in Si-compatible devices
have established acceptor states and relaxation
\cite{Song2011,vanDerHeijden2014}. &
Single-acceptor transport and spectroscopy have been reported
in Si-compatible devices, but no Ge acceptor-qubit array has
been demonstrated. \\

\midrule

\multicolumn{3}{@{}l}{
    \textbf{Gate-defined Ge holes---most mature Ge modality}
}\\
\addlinespace[0.3em]

One-qubit control &
An early two-qubit device demonstrated \(20~\mathrm{ns}\)
single-qubit gates with \(99.3\%\) fidelity. Later work reported
\(99.94\%\) fidelity at a low-field sweet spot, Rabi frequencies
above \(100~\mathrm{MHz}\), and subsequently above
\(500~\mathrm{MHz}\)
\cite{Hendrickx2020Fast,Wang2022,Hendrickx2024Sweet}. &
The values depend strongly on field orientation, confinement,
drive amplitude, and operating point. \\

Two-qubit control &
Two-qubit logic with a characteristic operation time of
\(75~\mathrm{ns}\) was reported in the 2020 benchmark device
\cite{Hendrickx2020Fast}. &
Exchange coupling is controlled electrically through the
interdot barrier and detuning gates. \\

Coherence and relaxation &
At a low-field sweet spot, representative values include
\(T_2^*=17.6~\mu\mathrm{s}\) and
\(T_2^{\mathrm{DD}}=1.3~\mathrm{ms}\)
\cite{Hendrickx2024Sweet}. &
Sweet-spot operation substantially reduces first-order
sensitivity to electrical noise. \\

Demonstrated scale &
{Four-qubit processor operation, coherent spin shuttling, and robust local control in a 10-spin array have been demonstrated. In addition, a 16-site Ge crossbar was tuned to odd hole occupation at every site and used to demonstrate shared-gate control of interdot tunnel coupling, establishing an important scaling milestone for dense two-dimensional arrays \cite{Borsoi2024Crossbar}. More recently, simultaneous initialization, single-qubit control, and readout were demonstrated in an 18-qubit modular Ge array, together with tunable exchange, controlled-Z operations, and three-qubit Greenberger--Horne--Zeilinger-state generation \cite{Dijkema2026Ge18Qubit}.} &
{The 16-site result establishes shared-control and array-tuning capability, whereas the 18-qubit result reports simultaneous qubit operation across the full modular array.} \\

\midrule

\multicolumn{3}{@{}l}{
    \textbf{Gate-defined Ge electrons---exploratory platform}
}\\
\addlinespace[0.3em]

One-qubit control &
No complete coherent-control benchmark has yet been reported
for an integrated gate-defined Ge electron-spin qubit. &
The spin-\(1/2\) encoding is conceptually simple, but electrical
control must manage valley, spin--orbit, and interface effects. \\

Two-qubit control &
No gate-defined Ge electron two-qubit gate has been
demonstrated. &
Electrically tunable exchange is the most direct proposed
interaction. \\

Coherence and relaxation &
A complete experimental set of \(T_1\), \(T_2^*\), echo
\(T_2\), and gate fidelity is not yet available. &
The platform remains at the quantum-dot and band-engineering
stage. \\

Assessment &
Gate-defined confinement offers post-fabrication tunability and
reconfigurability. &
These advantages are offset by the fourfold \(L\)-valley
structure, valley splitting, mass anisotropy, and sensitivity
to strain and interfaces. \\

\midrule

\multicolumn{3}{@{}l}{
    \textbf{Si electron-spin qubits---selected context}
}\\
\addlinespace[0.3em]

One-qubit control &
A representative one-qubit Clifford fidelity of up to
\(99.85\%\) has been reported above \(1~\mathrm{K}\)
\cite{Huang2024Above1K}. &
Included only as context; the Ge and Si material stacks,
devices, and benchmarking protocols differ. \\

Two-qubit control &
A two-qubit fidelity of \(98.92\%\) was reported at
\(1~\mathrm{K}\) \cite{Huang2024Above1K}. Separate Si MOS
devices achieved an average two-qubit fidelity of approximately
\(99.17\%\) at millikelvin temperature
\cite{Tanttu2024SiGates}. &
Multidot processors and industrially compatible MOS and SiGe
arrays have been demonstrated. \\

Coherence and relaxation &
The above-\(1~\mathrm{K}\) study reported a representative
\(T_1=12.23\pm2.11~\mathrm{ms}\)
\cite{Huang2024Above1K}. &
These values should not be used as a direct material-level
comparison with Ge because the operating conditions and device
architectures differ. \\

\midrule

\multicolumn{3}{@{}l}{
    {\textbf{Silicon hole-spin qubits---selected context}}
}\\
\addlinespace[0.3em]

Selected milestones &
{A single-hole qubit in natural-silicon MOS technology has exhibited an electrically tunable coherence sweet spot and Hahn-echo coherence approaching \(10^2~\mu\mathrm{s}\) \cite{Piot2022SiHole}. Two hole-spin qubits in a silicon FinFET have shown strongly anisotropic, electrically tunable exchange and fast conditional operations \cite{Geyer2024SiHoleExchange}. A recent foundry-platform preprint reports high-speed one- and two-qubit control with single-qubit fidelities approaching \(99.8\%\) and a two-qubit quality factor consistent with a high intrinsic fidelity ceiling \cite{Vorreiter2025SiHoleLogic}.} &
{These results show that hole-spin qubits are an active direction in both Ge and Si; Ge presently offers the more mature multiqubit hole platform, whereas Si holes benefit from direct compatibility with established silicon manufacturing.} \\

\end{xltabular}
\endgroup
\FloatBarrier

Silicon electron-spin qubits provide an important reference because isotopically enriched \(^{28}\)Si supports long coherence and a mature MOS/SiGe device toolbox. Recent Si devices report one- and two-qubit fidelities near or above \(99\%\), including operation above \(1~\mathrm{K}\), whereas gate-defined Ge holes offer exceptionally fast all-electrical control without micromagnets and have reached \(99.94\%\) one-qubit fidelity at a low-field sweet spot \cite{Huang2024Above1K,Tanttu2024SiGates,Hendrickx2024Sweet}. These numbers should not be compared as material constants: they reflect different magnetic fields, temperatures, encodings, pulse sets, and readout methods. The defensible conclusion is that Ge hole qubits are competitive in control speed and single-qubit performance, while Si retains greater breadth in reproducible electron-spin and multiqubit benchmarking.

Silicon hole-spin qubits provide a complementary comparison. Natural-silicon MOS, FinFET, and foundry-compatible devices now demonstrate coherence sweet spots, electrically tunable two-hole exchange, and high-speed control \cite{Piot2022SiHole,Geyer2024SiHoleExchange,Vorreiter2025SiHoleLogic}. These results reinforce that hole-spin qubits are developing in both material systems: Ge presently has the stronger multiqubit hole-qubit record, while Si offers direct integration with established manufacturing platforms.

The main lesson from Table~\ref{tab:quantitative_comparison} is that Ge does not support a single unified qubit story. Instead, it supports several distinct routes, each emphasizing a different compromise between microscopic simplicity, electrical controllability, fabrication tolerance, and systems-level scalability. Donor qubits are the most atom-like and conceptually closest to memory-oriented architectures. Acceptor qubits are the most native impurity realization of Ge valence-band spin-orbit physics and may ultimately be especially valuable in hybrid phononic or optical settings. Gate-defined hole qubits currently occupy the most advanced processor-oriented position in Ge, while gate-defined electron qubits remain scientifically interesting but still exploratory because they inherit the full complexity of the Ge L-valley conduction band without yet demonstrating a compensating systems-level advantage.

\subsection{Coherence}
If coherence is defined narrowly as the ability to isolate a spin from its environment, donor qubits have an intuitive advantage because they are more atom-like and less directly exposed to electrostatic disorder than gate-defined hole dots. In many semiconductor settings, one would therefore expect donors to offer the cleanest coherence ceiling. In Ge, however, this expectation is softened by the unusually strong role of spin-lattice relaxation. Once isotopic purification suppresses the host nuclear bath, donor-spin coherence is not freed indefinitely; instead it becomes limited by \(T_1\), reflecting the fact that strong spin-orbit coupling couples the spin more efficiently to phonons than in Si \cite{Sigillito2015}. Thus, while Ge donors retain important coherence advantages over more electrically active qubits, they do not inherit the same extraordinarily favorable hierarchy seen in the best Si donor systems.

Gate-defined hole qubits occupy the opposite end of the trade space. Their strong spin-orbit interaction makes them intrinsically more sensitive to electric fields and therefore, at first glance, more vulnerable to charge noise. Yet the experimental trajectory of the Ge hole platform has shown that this sensitivity can be mitigated rather than simply endured. Theory identified operating points where the qubit retains strong electrical controllability while suppressing first-order sensitivity to fluctuations \cite{Wang2021Optimal}, and experiment has since demonstrated sweet-spot operation with excellent coherence and very high fidelity \cite{Hendrickx2024Sweet}. Continued improvements in heterostructure quality, electrostatic design, and isotopic refinement further strengthen the case that Ge holes can be both fast and sufficiently coherent for scalable processors \cite{Stehouwer2025}.

Acceptor qubits remain harder to judge. In principle, they may eventually combine atom-like localization with strong electrical control and therefore occupy an appealing middle ground between donors and hole dots. However, in Ge this remains more of a theoretically motivated possibility than an experimentally benchmarked conclusion. Their coherence will likely depend sensitively on central-cell chemistry, interface position, strain environment, and dopant-specific hyperfine effects, all of which complicate any clean extrapolation from existing donor or hole results. Gate-defined electron qubits in Ge are similarly difficult to benchmark today: because the platform is still immature, there is no established coherence result competitive with the best hole data. For the moment, then, the coherence hierarchy is not simply ``atom-like is best''; rather, it is shaped by the distinct ways in which each Ge modality couples to phonons, fields, and local disorder.

\subsection{Control speed and power efficiency}
Gate-defined Ge holes currently set the pace in this category. Their strong spin-orbit interaction enables fast all-electrical manipulation through EDSR, avoiding the need for micromagnets or large local oscillating magnetic fields and thereby reducing device overhead and control complexity \cite{Watzinger2018,Wang2022,Hendrickx2020Fast}. This is not just a matter of convenience. Electrical driving with compact gate stacks is fundamentally attractive for scaling because it supports dense device layouts, localized actuation, and lower classical control overhead than architectures that rely more heavily on magnetic hardware.

Donor qubits in Ge can also exhibit substantial electrical tunability, especially through Stark shifts and hyperfine-assisted control. That makes them more electrically addressable than one might naively assume for an atom-like electron spin. Even so, their native control logic remains less directly all-electrical than that of Ge hole dots. In practical terms, this means that donor qubits may be well suited to architectures where tunable resonance and memory-like behavior are central, but they are not yet the natural benchmark for raw gate speed and local low-power control.

Acceptor qubits could, in principle, become formidable contenders in this category. Their spin-\(3/2\) structure, quadrupolar response, and strong coupling to electric fields and strain may eventually permit extremely efficient control, perhaps even rivaling conventional hole dots in certain regimes. However, that promise remains ahead of the Ge experimental base. Gate-defined electron qubits in Ge currently have the weakest case here. Because their spin degree of freedom is not as natively electric-field-active as that of holes, their route to fast all-electrical control is more indirect and would likely rely on engineered spin-orbit admixture or \(g\)-tensor modulation rather than on a naturally strong hole-like EDSR response. Until such control is demonstrated convincingly, Ge holes remain the clearest benchmark for fast and power-efficient operation within the Ge family.

\subsection{Scalability}
At the level of demonstrated scalability, gate-defined Ge holes are the least ambiguous case. The decisive point is not a single headline experiment, but the cumulative convergence of many milestones: single-spin operation, long relaxation times, fast two-qubit gates, a four-qubit processor, larger spin arrays, coherent shuttling, and continued improvement in wafer quality and device reproducibility \cite{Hendrickx2020Single,Hendrickx2021FourQubit,John2025,vanRiggelen2024,Stehouwer2025}. Taken together, these results show that the hole platform is no longer merely a promising physical qubit; it is becoming a serious architecture.

Donors, by contrast, are attractive for precision-oriented or memory-like designs, but they do not yet offer the same demonstrated scaling trajectory in Ge. Their atom-defined nature can be a strength at small scale, especially where electron--nuclear registers or highly reproducible local spectra are desired, yet atomic placement, activation control, and the absence of a mature large-circuit toolkit remain real constraints. Acceptors are even earlier in their architectural development. Their theoretical appeal for hybrid quantum technologies is strong, but the field is not yet at the point where one can speak credibly about acceptor-based processor scaling in Ge. Gate-defined electron qubits are earlier still: they have neither the experimental base of Ge holes nor the atomic identity advantage of donors.

For that reason, scalability is presently the clearest comparison category. If the target is a near-term multi-qubit Ge processor, gate-defined holes are the only modality with a substantial demonstrated lead. The other platforms may still become important in narrower or more specialized roles, but they are not yet competitive on this specific metric.

\subsection{Fabrication tolerance and reproducibility by confinement strategy}

Fabrication tolerance is most transparent when the four modalities are grouped according to how the carrier is confined. Impurity-defined donor and acceptor qubits require control of dopant species, isotope, depth, activation, and local crystal environment. Gate-defined electron and hole qubits avoid deterministic atomic placement, but transfer the reproducibility challenge to heterostructure quality, electrostatic disorder, gate geometry, and tunnel-barrier control.

\paragraph{Impurity-defined modalities.} Donor qubits benefit from atom-like confinement and a chemically well-defined central-cell potential, but their exchange and interface coupling depend strongly on donor position, valley composition, and distance from nearby electrodes or interfaces. Acceptor qubits share the placement challenge while adding a stronger sensitivity to local strain, electric-field gradients, and symmetry breaking of the spin-$3/2$ valence manifold. Consequently, acceptor devices place especially stringent demands on low-dislocation, low-strain-background material and reproducible interface geometry. For both impurity platforms, isotope selection is also part of device design because the dopant nuclear spin can be either a controlled register or an unwanted local noise source.

\paragraph{Gate-defined modalities.} Gate-defined dots allow the carrier position, occupancy, tunnel coupling, and exchange interaction to be tuned after fabrication. Ge hole dots presently provide the strongest experimental evidence that this electrostatic flexibility can support multiqubit operation, although their electrically tunable spin--orbit interaction also converts charge and strain disorder into qubit-frequency variability. Gate-defined Ge electron dots offer the same reconfigurability and a nominal spin-$1/2$ basis, but additionally require a sufficiently large and stable \(L\)-valley splitting. Their reproducibility therefore depends jointly on electrostatic uniformity and valley engineering at the interface.

The practical comparison is therefore not between “precise” impurity devices and “tolerant” gate-defined devices.
Each confinement strategy moves the dominant tolerance requirement to a different physical layer: atomic placement and local symmetry for impurity qubits, versus heterostructure, interface, and gate-stack uniformity for electrostatic dots. At present, gate-defined Ge hole qubits offer the clearest evidence that these requirements can be managed across multiqubit devices.

\subsection{Which modality is best for what?}
The answer depends on the scientific or technological objective rather than on any single metric.

For a \textbf{near-term scalable processor tile}, gate-defined Ge holes are the strongest choice. They have the best experimental base, the clearest scaling path, and the most mature control toolbox. Their combination of all-electrical manipulation, compact gate layouts, and demonstrated multi-qubit operation makes them the natural processor platform in Ge \cite{Hendrickx2020Fast,Hendrickx2021FourQubit,John2025}.

For an \textbf{atom-like quantum memory or hybrid electron--nuclear register}, donor qubits deserve continued serious attention. Their microscopic reproducibility, hybrid electron--nuclear structure, and strong Stark tunability make them especially attractive where memory, local spectral addressability, or atom-defined functionality matters more than rapid expansion to dense processor arrays \cite{Sigillito2015,Sigillito2016Stark}.

For \textbf{phononic, photonic, or quadrupolar hybrid architectures}, acceptor qubits may ultimately prove exceptionally valuable. Their spin-\(3/2\) manifold and strong electric- and strain-response give them a distinctive role that neither donors nor conventional gate-defined dots replicate naturally. However, realizing that promise in Ge will require a stronger experimental foundation than currently exists \cite{AbadilloUriel2016,Salfi2016ChargeInsensitive,AbadilloUriel2023StrainSOI}.

For a \textbf{spin-\(1/2\) Ge quantum-dot route} analogous to Si electrons, gate-defined electrons remain a longer-range possibility rather than a current contender. Their conceptual simplicity is appealing, but they still lack both the maturity and the demonstrated system-level advantages needed to compete seriously with Ge holes at present \cite{Baron2003Lvalleyg,Giorgioni2016GeQW,Virgilio2009GeValley}.

The overall conclusion is therefore not that one modality should replace all others. Rather, Ge appears to support a layered ecosystem of qubit types. For processor-oriented applications, gate-defined holes presently offer the strongest combination of demonstrated control, scalability, and architectural readiness. Donors remain important for atom-like and memory-centered functions; acceptors are strategically interesting for hybrid spin--phonon and spin--photon directions; and gate-defined electrons remain exploratory.

\section{Roadmap and scientific priorities}

A useful comparison should end not with a ranking alone, but with a roadmap. The central lesson of the preceding sections is that the four Ge-based qubit modalities do not sit on the same developmental curve. Gate-defined hole qubits in strained Ge/SiGe are already in an architecture-building phase, where the main questions concern array scaling, calibration robustness, and fault-tolerant compatibility. Donor qubits are at an earlier but still strategically important stage, where the path forward is clear if the materials and two-qubit milestones can be met. Acceptor qubits remain the most scientifically intriguing impurity-based route for hybrid spin--phonon and spin--photon concepts, but they still require basic closure on species selection, spectroscopy, and reproducibility. Gate-defined electron qubits in Ge are further behind and must first justify their strategic role relative to the already successful hole platform. For that reason, the roadmap should not ask every modality to solve the same next problem. Instead, each platform should be advanced according to its own dominant bottleneck.

\subsection{Priority directions for donor qubits}
The donor route needs three linked advances. The first is a materials advance: isotopically purified Ge host crystals with carefully controlled residual \(^{73}\)Ge and sufficiently low residual chemical disorder to expose the intrinsic donor-spin limit in a reproducible way \cite{Sigillito2015}. Without that foundation, it is difficult to separate true donor physics from sample-specific broadening and relaxation. The second is a device-fabrication advance: deterministic donor placement, activation, and interface registration at a level consistent with controlled exchange or other tunable two-qubit interactions \cite{Pica2016}. Ge donors are in principle somewhat more forgiving than Si donors because of their larger effective Bohr radii, but that advantage only matters if the nanofabrication stack can actually exploit it. The third is an architectural advance: direct demonstrations of entangling gates, exchange control, or hybrid electron--nuclear logic in Ge, rather than continued reliance on coherence and Stark-shift characterization alone \cite{Sigillito2016Stark}.

A particularly important intermediate goal is to establish whether Ge donors can occupy a distinctive role rather than merely reproduce a weaker version of the Si donor program. The most promising niche is likely not a dense processor tile, but a hybrid architecture in which the electron provides tunable control and readout while the nuclear degree of freedom serves as a longer-lived memory. In that context, large Stark tunability, atom-defined confinement, and species-specific hyperfine structure become genuine system-level assets rather than mere scientific curiosities. A second promising direction is hybrid donor--phononic integration, where engineered acoustic environments could be used either to suppress unwanted spin-lattice relaxation channels or to couple selected donors to localized modes. The key near-term question is therefore not simply ``can Ge donors be coherent,'' because that has already been answered qualitatively, but rather ``can Ge donors be made architecturally useful in a way that exploits what is special about Ge?''

\subsection{Priority directions for acceptor qubits}
For acceptors, the first task is not immediately high-fidelity logic but \emph{materials and spectroscopy closure}. This platform is still at the stage where basic questions remain open. Which acceptor species in Ge is most favorable when one weighs central-cell physics, hyperfine burden, activation behavior, and interface compatibility? How reproducibly can the interface-split Kramers manifold be engineered across nominally identical devices? What hyperfine floor is imposed by the dopant nucleus even when the host Ge is isotopically improved? Can one reliably define sweet spots or protected operating regimes in which electrical control remains strong while first-order noise sensitivity is reduced \cite{AbadilloUriel2016,Salfi2016ChargeInsensitive,AbadilloUriel2018MagicAngles}? Until these questions are answered experimentally, acceptor qubits in Ge remain architecturally intriguing but immature.

A second priority is to move from theory-rich intuition to experimentally benchmarked device physics. That means reproducible single-acceptor spectroscopy, direct measurements of electric-field control at the level of a single acceptor, and a systematic map of how interface depth, local strain, and dielectric environment reshape the effective qubit manifold \cite{AbadilloUriel2016,AbadilloUriel2023StrainSOI}. Only after that closure is obtained does it make sense to ask whether acceptors should be optimized for conventional quantum logic, for protected electrically controlled qubits, or for hybrid coupling to phonons, cavities, and optical channels. In fact, the most compelling long-term case for Ge acceptors may lie precisely in those hybrid directions. Their spin-\(3/2\) structure, quadrupolar response, and strain sensitivity are unusual strengths, but they will only become credible architectural assets once the field has demonstrated that the relevant manifolds can be prepared, tuned, and read out reproducibly from device to device.

\subsection{Priority directions for hole qubits}
For Ge holes the agenda is more advanced and more concrete. Here the goal is not to discover whether the platform works, but to convert an already successful qubit technology into a robust large-scale architecture. The first priority is continued improvement of the host and gate stack: full isotopic purification where beneficial, lower dielectric defect density, reduced charge noise, and more uniform electrostatics across wafers and across large arrays \cite{Hendrickx2024Sweet,Stehouwer2025}. The second is better control of the electrically tunable \(g\)-tensor and its anisotropy. This is one of the great strengths of the Ge hole platform, but it also complicates calibration because qubit frequencies, Rabi rates, and noise sensitivity depend strongly on bias and field orientation \cite{Wang2021Optimal,Hendrickx2024Sweet}. In small devices this is manageable; in larger arrays it becomes a major systems-engineering problem.

The third priority is architectural scaling: robust tuning protocols for multi-qubit arrays, high-fidelity two-qubit gates in geometries compatible with error-correcting layouts, and continued progress in shuttling, connectivity, and modular control \cite{Hendrickx2020Fast,Hendrickx2021FourQubit,John2025,vanRiggelen2024}. Recent results strongly suggest that these are tractable engineering directions rather than unknown physics. The key unknown is no longer whether Ge holes can be coherent and controllable, but how efficiently the platform can be industrialized into arrays with acceptable calibration overhead, reproducibility, and control complexity. A fourth, slightly longer-range direction is selective hybridization: once the base hole-qubit hardware is sufficiently stable, integrating phononic crystal cavities, resonators, or other transduction elements becomes scientifically attractive because the same spin-orbit and strain response that enables fast EDSR control may also support controlled spin--phonon coupling \cite{Mauro2025}. In other words, for Ge holes the roadmap naturally separates into a near-term processor track and a longer-term hybrid-quantum track.

\subsection{Priority directions for gate-defined electrons}
The main question for Ge electrons is whether they can reach a regime where their simpler spin-\(1/2\) logic outweighs the complications of L-valley physics. If not, they will remain scientifically interesting but strategically secondary to Ge holes. The first priority is therefore not processor scaling, but basic stack validation: can strain-engineered Ge/SiGe heterostructures support a sufficiently clean and reproducible electron channel with controllable valley splitting, reproducible few-electron confinement, and an experimentally useful \(g\)-tensor landscape \cite{Baron2003Lvalleyg,Giorgioni2016GeQW,Virgilio2009GeValley}? Without those elements, there is no meaningful electron-qubit roadmap to discuss.

The second priority is to establish whether Ge electrons can offer a real control or coherence advantage in any operating regime. That requires single-electron loading, spin initialization and readout, direct measurements of spin relaxation and dephasing, and some demonstration of electrically assisted control that is competitive enough to justify continued investment. If the electron route can only reproduce a weaker and less mature version of what Ge holes already achieve, then its role will remain exploratory. If, however, L-valley engineering, confinement-driven \(g\)-tensor modulation, or a hybrid spin--phonon design reveals a distinctive operating point unavailable to holes, then the platform may justify a more serious long-term effort. The roadmap for Ge electrons is therefore a decision tree rather than a straightforward scaling program: first prove the stack, then prove a distinctive advantage, and only after that ask whether scaling is worthwhile.

Taken together, these priorities suggest a layered Ge quantum roadmap rather than a winner-take-all strategy. Gate-defined Ge holes should continue to receive strong attention for processor-oriented development because they currently offer the clearest combination of experimental maturity and scalability. Donors deserve focused investment as a complementary atom-like and memory-capable modality, especially if deterministic placement and two-qubit logic can be demonstrated. Acceptors should be pursued as a strategically important hybrid platform, but with the clear recognition that the immediate challenge is still spectroscopy and reproducibility rather than processor performance. Gate-defined electrons should remain an exploratory effort until they demonstrate a regime in which their conceptual simplicity yields a practical advantage over holes. This is a healthy scientific landscape: not one modality replacing the others, but several routes developing in parallel, each justified by a different combination of physics opportunity and architectural promise.

\section{Conclusions}

The phrase ``germanium qubit'' does not describe a single technology. It describes a family of physically distinct spin-qubit platforms that share a common semiconductor host but rely on different confinement mechanisms, band-edge physics, control channels, and scaling strategies. This distinction is central to any realistic assessment of Ge-based quantum hardware. Donor qubits, acceptor qubits, gate-defined hole qubits, and gate-defined electron qubits all benefit from the materials advantages of Ge, including isotopic purification, high mobility, mature processing, and access to very high chemical purity. However, they use these advantages in different ways and face different limiting mechanisms.

Donor qubits in Ge offer atom-defined confinement, large Stark tunability, and a natural route to hybrid electron--nuclear registers. These features make them attractive for memory-oriented architectures, frequency-addressable arrays, and hybrid quantum systems. At the same time, Ge donor spins are more strongly constrained by spin-lattice relaxation than their Si counterparts because the stronger spin--orbit interaction and multivalley conduction-band structure enhance phonon-mediated relaxation. Their practical development therefore depends not only on deterministic donor placement and activation, but also on engineering the phonon environment, demonstrating robust two-qubit coupling, and realizing hybrid electron--nuclear functionality in actual Ge devices \cite{Sigillito2015,Sigillito2016Stark,Pica2016}.

Acceptor qubits represent the most direct impurity-based realization of Ge valence-band spin--orbit physics. Their spin-\(3/2\) structure, quadrupolar response, and strong coupling to electric fields and strain make them especially appealing for electrically driven, strain-coupled, and cavity-enabled architectures. In principle, they occupy an attractive middle ground between donor qubits and gate-defined hole qubits: they retain chemically defined localization while inheriting the rich control physics of the Ge valence band. In practice, however, Ge acceptor qubits remain experimentally immature. The immediate challenges are still basic spectroscopy, impurity-species selection, interface and strain control, reproducible single-acceptor operation, and the demonstration of reliable readout and coupling, rather than processor-level logic \cite{AbadilloUriel2016,Salfi2016ChargeInsensitive,AbadilloUriel2023StrainSOI}.

Gate-defined electron qubits in Ge preserve the conceptual simplicity of spin-\(1/2\) quantum-dot qubits and could, in principle, benefit from the extensive control logic developed for Si electron-spin devices. Their attraction is therefore clear: they offer a familiar spin encoding while remaining compatible with electrostatic confinement in Ge/SiGe heterostructures. Their main difficulty is that they inherit the full complexity of the Ge conduction band, including the fourfold \(L\)-valley structure, anisotropic effective masses, valley-orbit coupling, and strong sensitivity to confinement, interfaces, and strain. At present, gate-defined Ge electron qubits have not yet demonstrated a compensating experimental advantage that would place them on the same footing as the more advanced Ge hole-spin platform. They should therefore be viewed as a scientifically interesting but still exploratory direction. Their long-term value will depend first on achieving a sufficiently large, stable, and reproducible valley splitting, together with quantified control of intervalley mixing and leakage \cite{Baron2003Lvalleyg,Giorgioni2016GeQW,Virgilio2009GeValley}.

Among the four modalities, gate-defined hole spin qubits in Ge nanostructures and strained Ge/SiGe heterostructures have advanced furthest toward architecture-relevant operation. They have demonstrated the broadest set of experimentally mature capabilities in the present Ge ecosystem, including single-spin control, single-shot readout, fast two-qubit logic, multi-qubit processor operation, coherent spin shuttling, and continued progress toward low-noise, array-compatible device designs \cite{Watzinger2018,Hendrickx2020Fast,Hendrickx2024Sweet}. Their strength lies in the combination of valley-free valence-band physics, strong and electrically tunable spin--orbit coupling, light in-plane effective mass, and compatibility with planar semiconductor fabrication. The main challenges for Ge hole qubits are therefore no longer questions of basic feasibility, but questions of scale: suppressing charge noise, improving wafer and gate-stack uniformity, managing \(g\)-tensor anisotropy, maintaining tunability across larger arrays, and developing robust control strategies compatible with cryogenic integration.

A technically balanced conclusion is therefore not that one Ge modality makes the others irrelevant. Rather, Ge supports a layered qubit ecosystem in which different platforms may serve different roles. For processor-oriented applications, gate-defined hole qubits presently offer the strongest combination of experimental maturity, electrical controllability, and architectural scalability. Donor qubits remain important as atom-like, highly tunable systems with potential value for memory-oriented, hybrid electron--nuclear, and phonon-mediated architectures. Acceptor qubits provide a compelling but earlier-stage route to impurity-defined valence-band qubits, especially for future spin--phonon, spin--photon, and cavity-based devices. Gate-defined electron qubits remain a longer-range possibility whose future impact will depend on whether the complexity of the Ge \(L\)-valley conduction band can be controlled sufficiently for reproducible qubit operation.

Phononic-crystal engineering emerges from this comparison as a shared but modality-dependent design axis. For electron-like systems, including donor and gate-defined electron qubits, PnC structures primarily offer a way to suppress phonon-mediated spin-lattice relaxation and to enable virtual-phonon or cavity-assisted coupling. For hole-like systems, including acceptor and gate-defined hole qubits, the role of PnC engineering can be broader: because strain couples directly to the spin--orbit-active valence-band manifold, phononic structures can serve not only as relaxation filters but also as active control, transduction, and coupling elements. In this sense, PnC geometry should be viewed as a cross-platform architectural tool for Ge spin qubits, but one whose optimal use depends strongly on whether the qubit is electron-like or hole-like.

The broader message is that the Ge qubit landscape should not be interpreted as a single race with one universal winner. Instead, it is a set of parallel and complementary directions enabled by the same unusually rich semiconductor host. The near-term path to scalable Ge quantum processors is most clearly led by gate-defined hole-spin qubits. The longer-term opportunity is broader: high-purity and isotopically engineered Ge may support a family of qubit, memory, sensing, and phononic-coupling technologies in which donor, acceptor, electron, and hole platforms occupy different roles within an integrated quantum-technology stack.

\section*{Acknowledgment}
This work was supported in part by NSF OISE 1743790, NSF PHYS 2117774, NSF OIA 2427805, NSF PHYS 2310027, NSF OIA 2437416, DOE DE-SC0024519, DE-SC0004768, and a research center supported by the State of South Dakota. 

\bibliographystyle{unsrt}
\bibliography{ge_qubits_comparison}

\end{document}